\documentclass[12pt]{article}
\usepackage[utf8]{inputenc}
\usepackage{amsmath}

\usepackage{slashed}
\usepackage[table]{xcolor}
\usepackage{xcolor}
\usepackage{blindtext}
\usepackage{multicol}
\usepackage{graphicx}
\usepackage{verbatim}
\usepackage{amssymb}
\usepackage{adjustbox}
\usepackage{hyperref}
\usepackage{multirow}
\usepackage{tikz-feynman}
\tikzfeynmanset{compat=1.0.0}
\usepackage{setspace}
\usepackage[margin=60pt]{geometry}
\usepackage{graphicx,caption}

\newcommand{\be}{\begin{equation}}
\newcommand{\ee}{\end{equation}}
\newcommand{\ba}{\begin{eqnarray}}
\newcommand{\ea}{\end{eqnarray}}
\newcommand{\bs}{\begin{subequations}}
\newcommand{\es}{\end{subequations}}
\newcommand{\no}{\nonumber\\}

\title{\huge Unitarity constraints on large multiplets
  of arbitrary gauge groups}
\author{\\
  Andr\'e Milagre\footnote{E-mail: andre.milagre@gmail.com}
  \ and\ Lu\'\i s Lavoura\footnote{E-mail: balio@cftp.tecnico.ulisboa.pt}
\\*[3mm]
\small Universidade de Lisboa, Instituto Superior T\'ecnico, CFTP, \\
\small Av.~Rovisco Pais~1, 1049-001 Lisboa, Portugal
\vspace*{7mm}
}

\date{\today}

\begin{document}

\maketitle

\begin{abstract}
  We impose partial-wave unitarity on $2 \to 2$ tree-level
  scattering processes to derive constraints
  on the dimensions of large scalar and fermionic multiplets
  of arbitrary gauge groups.
  We apply our results to scalar and fermionic extensions
  of the Standard Model,
  and also to the Grand Unified Theories (GUTs)
  based on the groups $SU(5)$,
  $SO(10)$,
  and $E_6$.
  We find scenarios within the latter two GUTs
  that violate the unitarity condition;
  this may require a reevaluation
  of the validity of perturbation theory in those scenarios.
\end{abstract}

\section{Introduction}

The conservation of probability in quantum-mechanical processes implies unitarity of the $S$ matrix. When this condition is phrased in terms of the
partial waves
of a scattering amplitude, one may derive constraints on the magnitude of the zeroth
partial wave,
if one wants perturbation theory to be trusted in the theory in question. Historically, a famous application of partial-wave unitarity bounds was made by Lee, Quigg, and Thacker~\cite{Lee:1977yc, Lee:1977eg} to derive the upper bound $m_H \lesssim 1\,\textup{TeV}$ on the mass $m_H$ of the Higgs boson. The same reasoning has been applied by Logan and her collaborators~\cite{Hally:2012pu} to constrain the quantum numbers of large scalar multiplets in extensions of the Standard Model (SM); they have shown that one complex scalar multiplet with more than eight components (or one real multiplet with more than nine components) is enough to violate perturbation theory. They also found constraints on the hypercharges of such multiplets.
More recently, by working with fields transforming in the trivial, fundamental, 
and adjoint representations of $SU(N) \times U(1)$, upper bounds on Yukawa \cite{Allwicher:2021rtd} and gauge \cite{Barducci:2023lqx} couplings have been derived by using perturbative unitarity.
One-loop corrections to the Higgs--$WW$ and Higgs--$ZZ$ couplings have also been used to constrain the dimension of large electroweak representations of vector-like fermions \cite{DAgnolo:2023rnh}.

The goal of this work is to extend Ref.~\cite{Hally:2012pu}
by using the unitarity bounds on the zeroth
partial wave
of $2 \to 2$ tree-level scattering amplitudes
to constrain the dimensions of large scalar \emph{and fermionic} multiplets
that transform under \emph{arbitrary} gauge groups.
We compute both the fermion-pair and the scalar-pair annihilation
into gauge-boson pairs in the high-energy limit.
The calculations are performed in the unmixed basis of each gauge group
and before spontaneous symmetry breaking.
Therefore the scalars,
the fermions,
and the gauge bosons are taken to be massless,
and the latter only have transverse polarizations.

In Section~\ref{sec:sec1} we provide a brief review
of the unitarity constraints on the zeroth
partial wave
of $2 \to 2$ tree-level scattering amplitudes;
we then explicitly calculate the zeroth
partial wave
for a single
fermion pair or scalar pair annihilating
into a gauge-boson pair of a single gauge group.
In Section~\ref{sec:sec2} we introduce the technique
of coupled-channel analysis to constrain the dimension
of a single fermion or scalar multiplet under a single gauge group;
we then generalize the reasoning to multiple multiplets
transforming under multiple gauge groups.
We apply our results to derive constraints
on extensions of the SM and on the $SU(5)$,
$SO(10)$,
and $E_6$ Grand Unified Theories (GUTs) in Section~\ref{sec:sec3}, 
and draw our conclusions in Section \ref{sec:conclusion}.
We use Appendix \ref{sec:appA} to carefully derive the scattering amplitudes used in Section \ref{sec:sec2}.
In Appendix \ref{appen:a0_example2} we substantiate an assumption made about the largest eigenvalue of the coupled-channel matrix,
by using a pedagogical example. 
In Appendix~\ref{appen:group_theory}
we provide useful properties for the relevant irreducible representations of the various gauge groups
used in this paper.

\section{$2 \to 2$ scatterings and unitarity constraints}
\label{sec:sec1}

\subsection{Interactions of the scalars and fermions with the gauge bosons}

We consider a quantum field theory symmetric under a Lie Group $\mathcal{G}$. In this theory, a fermion multiplet is placed in a representation $\textbf{\textit{n}}_\psi$ of $\mathcal{G}$, while a scalar multiplet is placed in a representation $\textbf{\textit{n}}_\varphi$ of $\mathcal{G}$:
\begin{align}
\textbf{\textit{n}}_\psi &= 
\begin{pmatrix}
\psi_1\\ 
\psi_2\\ 
\vdots \\ 
\psi_n
\end{pmatrix},	&
\textbf{\textit{n}}_\varphi = 
\begin{pmatrix}
\varphi_1\\ 
\varphi_2\\ 
\vdots \\ 
\varphi_n
\end{pmatrix}.
\end{align}
Without loss of generality,
we let $n$ denote both the dimension of $\textbf{\textit{n}}_\psi$ 
and the one of $\textbf{\textit{n}}_\varphi$.

The covariant derivative acting on either multiplet
is defined as\footnote{A sum over repeated indices
is always implied in this paper.}~\cite{Romao:2012pq}
\bs
\ba
D_\mu \textbf{\textit{n}}_\psi &=&
\partial_\mu \textbf{\textit{n}}_\psi
+ i g V^a_\mu\, T_a \textbf{\textit{n}}_\psi,	\\
D_\mu \textbf{\textit{n}}_\varphi
&=& \partial_\mu \textbf{\textit{n}}_\varphi
+ i g V^a_\mu\, T_a \textbf{\textit{n}}_\varphi.
\ea
\es
The $T_a$ are the $n \times n$ matrices that represent the generators of $\mathcal{G}$; the $V_\mu^a$ are $\tilde{n}$ gauge bosons which transform in the $\tilde{n}$-dimensional adjoint representation of $\mathcal{G}$; $g$ is the gauge coupling constant.

The interactions of the fermions with the gauge bosons
derive from
\bs
\ba
\mathcal{L} &\supset&
i \, \overline{\textbf{\textit{n}} }_\psi \slashed{D}   \textbf{\textit{n}}_\psi
\\ &=&
i \, \overline{\psi}_i    \slashed{\partial}   \psi_i
- g \left( T_a \right)_{ji} \overline{\psi }_j\,   \slashed{V}^a   \psi_i.
\ea
\es
The scalars interact with gauge bosons through
\bs
\label{pio}
\ba
\mathcal{L} &\supset&
D_\mu \textbf{\textit{n}}_\varphi^\dagger\, D^\mu \textbf{\textit{n}}_\varphi
\\ &=&
\left( \partial_\mu  \varphi_i \right)^\dagger
\left( \partial^\mu  \varphi_i \right)
+ ig \left( T_a \right)_{ji} V_\mu^a
\left(  \varphi_i\, \partial^\mu \varphi_j^\dagger
-   \varphi_j^\dagger\,  \partial^\mu \varphi_i \right)
+ g^2 \left( T_a T_b \right)_{ji}
V_\mu^a\, V^{\mu, b}\,  \varphi_j^\dagger  \,  \varphi_i.
\ea
\es
If $\textbf{\textit{n}}_\varphi$ is in a real representation of $\mathcal{G}$,
then one must drop the $\dagger$'s from Eq.~\eqref{pio}
and add a factor $1/2$ to its right-hand side.

\subsection{Unitarity bounds}
\label{sec:uni_bounds}

We want to consider scattering processes of the type $\psi_i \overline{\psi}_j \to V^a V^b$ or $\varphi_i \varphi^\dagger_j \to V^a V^b$. The amplitude of each process may be decomposed into partial-wave amplitudes according to the formula~\cite{Schwartz:2014sze}
\begin{equation}
  \mathcal{M} \left( \cos \theta \right) = 16 \pi \sum_{J=0}^\infty a_J
  \left( 2J+1 \right) P_J \left( \cos{\theta} \right),
\end{equation}
where $\theta$ is the scattering angle and
$J$ (which is a non-negative integer)
is the total orbital angular momentum of the final state. The $a_J$ are numerical coefficients. The functions $P_J \left( \cos{\theta} \right)$ are the Legendre polynomials; in particular,
\begin{equation}
  P_0 \left( \cos{\theta} \right) = 1.
\end{equation}
The Legendre polynomials satisfy the orthogonality relation
\begin{equation}
  \int_{-1}^{+1} \! P_J \left( \cos{\theta} \right)
  P_{J'} \left( \cos{\theta} \right) \,
  \mathrm{d} \cos{\theta} = \frac{2}{2J+1}\, \delta_{J J'}.
\end{equation}
Therefore, the zeroth
partial wave is given by
\begin{equation}
  \label{aaa000}
  a_0 \equiv \frac{1}{32 \pi}\, \int_{-1}^{+1}
  \mathcal{M} \left( \cos{\theta} \right) \mathrm{d} \cos\theta.
\end{equation}

Tree-level zeroth partial-wave unitarity dictates
that~\cite{Schwartz:2014sze, Logan:2022uus}:
\bs
\ba
\left| \mathrm{Re}\, a_0 \right| &\le& \frac{1}{2},
\label{unitarity_condition}
\\
\left| a_0 \right| &\le& 1.
\ea
\es
In this work, we want to constrain the dimension of large fermionic or scalar multiplets by using condition~\eqref{unitarity_condition}. In order to find the most stringent constraint we employ the technique of coupled-channel analysis. The zeroth
partial waves
will be calculated for each process $\psi_i \overline{\psi}_j \to V^a V^b$ and/or $\varphi_i \varphi^\dagger_j \to V^a V^b$, and will then be assembled in a coupled-channel matrix
for various pair of indices $\left( i, j \right)$ and $\left( a, b \right)$.
The unitarity bound arises from applying condition~\eqref{unitarity_condition}
to the largest (in modulus) eigenvalue of the coupled-channel matrix~\cite{Logan:2022uus, Ginzburg:2005dt, Kanemura:2015ska}. It should be pointed out that the
partial wave
for a process involving two identical initial-state or final-state particles receives an extra factor $1 \left/ \sqrt{2} \right.$;
if the two initial-state particles are identical
and the two final-state particles are also identical,
then the extra factor is $1/2$.

\subsection{Fermionic Scattering}

We want to calculate the zeroth
partial wave of the process
\begin{equation}
	\psi_i\, \overline{\psi}_j \to V^a V^b.
\end{equation}

\noindent Three channels contribute to the corresponding amplitude:

\begin{equation}
i\mathcal{M}(ij\to ab) \ =
\  \begin{tikzpicture}[baseline={([yshift=-1pt]current bounding box.center)},vertex/.style={anchor=base,
    circle,fill=black!25,minimum size=18pt,inner sep=2pt}]
\begin{feynman}
	\vertex (a1) at (-1.7,1.) {$\psi_{i}$};
	\vertex (a2) at (0,0.8) ;
	\vertex (a3) at (1.7,1.) {$V^a_\mu$};
	\vertex (b1) at (-1.7,-1.) {$\overline{\psi}_{j}$};
	\vertex (b2) at (0,-0.8) ;
	\vertex (b3) at (1.7,-1.) {$V^b_\nu$};

	\diagram*{ (a1) -- [fermion, momentum=\( p_1 \)](a2),
	(a2) -- [photon, momentum=\( p_3 \)](a3),
	(a2) -- [fermion, edge label=\( \psi_k \)](b2),
	(b1) -- [anti fermion, momentum'=\( p_2 \)](b2),
	(b2) -- [photon, momentum'=\( p_4 \)](b3),};
\end{feynman}
\end{tikzpicture}\ \ +\ \ 
\begin{tikzpicture}[baseline={([yshift=-1pt]current bounding box.center)},vertex/.style={anchor=base,
    circle,fill=black!25,minimum size=18pt,inner sep=2pt}]
\begin{feynman}
	\vertex (a1) at (-1.7,1.) {$\psi_{i}$};
	\vertex (a2) at (0,0.8) ;
	\vertex (a3) at (1.7,1.) {$V^b_\nu$};
	\vertex (b1) at (-1.7,-1.) {$\overline{\psi}_{j}$};
	\vertex (b2) at (0,-0.8) ;
	\vertex (b3) at (1.7,-1.) {$V^a_\mu$};

	\diagram*{ (a1) -- [fermion, momentum=\( p_1 \)](a2),
	(a2) -- [photon, momentum=\( p_4 \)](a3),
	(a2) -- [fermion, edge label=\( \psi_k \)](b2),
	(b1) -- [anti fermion, momentum'=\( p_2 \)](b2),
	(b2) -- [photon, momentum'=\( p_3 \)](b3),};
\end{feynman}
\end{tikzpicture}\ \ +\ \ 
\begin{tikzpicture}[baseline={([yshift=-1pt]current bounding box.center)},vertex/.style={anchor=base,
    circle,fill=black!25,minimum size=18pt,inner sep=2pt}]
\begin{feynman}
	\vertex (a1) at (-1.7,1.1) {$\psi_{i}$};
	\vertex (a2) at (-0.9,0) ;
	\vertex (a3) at (1.7,1.1) {$V_\mu^a$};
	\vertex (b1) at (-1.7,-1.1) {$\overline{\psi}_{j}$};
	\vertex (b2) at (0.9,0) ;
	\vertex (b3) at (1.7,-1.1) {$V_\nu^b$};

	\diagram*{ (a1) -- [fermion, momentum=\( p_1 \)](a2),
	(b1) -- [anti fermion, momentum'=\( p_2 \)](a2),
	(a2) -- [photon, edge label=\( V^c \)](b2),
	(b2) -- [photon, momentum=\( p_3 \)](a3),
	(b2) -- [photon, momentum'=\( p_4 \)](b3),};
\end{feynman}
\end{tikzpicture}.
\label{diags_fermion}
\end{equation}

\vspace{5pt}

\noindent We work in the high-energy limit and we therefore neglect
the masses of the fermions and gauge bosons.
In that limit, fermions become chiral, meaning
$\overline{\psi}_j$
has opposite chirality to $\psi_i$.
Consequently,
if the two final-state gauge bosons have opposite polarizations
(either left--right or right--left),
then the sum of the three diagrams in Eq.~\eqref{diags_fermion} vanishes.
On the other hand,
if the two final-state gauge bosons have the same polarization
(either left--left or right--right),
then the third diagram (the $s$ channel) in Eq.~\eqref{diags_fermion} vanishes
and only the first two are nonzero.
The amplitudes for the $t$- and $u$-channels read
\begin{subequations}
\begin{align}
\mathcal{M}_t \left( ij \to ab \right) &=
  -g^2 \left( T_b T_a \right)_{ji}\,
  \frac{\overline{v}_{j}\, \slashed{\varepsilon}^* \left( p_4 \right)
      \left( \slashed{p}_1 - \slashed{p}_3 \right)
      \slashed{\varepsilon}^* \left( p_3 \right) u_i}{t},
\label{ff_tchan}\\
\mathcal{M}_u(ij\to ab) &= -g^2 \left( T_a T_b\right)_{ji}\,
\frac{\overline{v}_{j}\, \slashed{\varepsilon}^* \left( p_3 \right)
    \left( \slashed{p}_1 - \slashed{p}_4 \right)
    \slashed{\varepsilon}^* \left( p_4 \right) u_i}{u},
\label{ff_uchan}
\end{align}
\end{subequations}
respectively,
where 
\begin{description}
\item $u_i$ and $\overline{v}_j$ are the the spinors
  of the initial-state $\psi_i$ and $\overline{\psi}_j$,
  respectively,
\item $\varepsilon \left( p_3 \right)$
  and $\varepsilon \left( p_4 \right)$ are the polarization four-vectors
  of the final-state $V^a$ and $V^b$, respectively,
\item $t \equiv \left( p_1 - p_3 \right)^2$
  and $u \equiv \left( p_1 - p_4 \right)^2$ are the Mandelstam variables.
\end{description}

Since the gauge bosons have the same polarization, either $V^a$
and $V^b$ are both right-handed (RR configuration),
or they are both left-handed (LL configuration).
Using the results of Appendix~\ref{sec:fermion_explicit_calculation}, the total amplitude for each scenario is found to be
\bs
\ba
\mathcal{M}_{\textup{RR}}(ij\to ab) &=& -g^2  \sin \theta \left [  \left (T_a T_b  \right )_{ji}   \tan^2 \frac{\theta}{2}  +  \left (T_b T_a  \right )_{ji} \right ],\\
\mathcal{M}_{\textup{LL}}(ij\to ab) &=& g^2  \sin \theta \left [  \left (T_a T_b  \right )_{ji}    +  \left (T_b T_a  \right )_{ji}  \cot^2 \frac{\theta}{2} \right ].
\ea
\es

\noindent We use Eq.~\eqref{aaa000} for $a_0$ and\footnote{Notice
the factors $\pi$ in the integrals~\eqref{jvppp}.
Those factors do not appear in the corresponding integrals
for $\varphi_i \varphi_j^\dagger \to V^a V^b$.
They are responsible for a factor $\pi / 2$
in the comparison of the $a_0$'s
for fermionic-pair and scalar-pair scattering to gauge-boson pair.}
\bs
\label{jvppp}
\ba
\int_{-1}^{+1} \sin{\theta}\, \tan^2{\frac{\theta}{2}}\ \mathrm{d} \cos{\theta} 
=\int_{-1}^{+1} \sin{\theta}\, \cot^2{\frac{\theta}{2}}\ \mathrm{d} \cos{\theta}
&=& \frac{3 \pi}{2},
\\
\int_{-1}^{+1} \sin{\theta}\ \mathrm{d} \cos{\theta}
&=& \frac{\pi}{2},
\ea
\es
to obtain
\begin{subequations} \label{uti} \begin{eqnarray}
    a_0^{\textup{RR}} \left( ij\to ab \right) &=&
    - \frac{g^2}{64} \left( 3 T_a T_b + T_b T_a \right)_{ji},
    \label{iuprd} \\
    a_0^{\textup{LL}} \left( ij\to ab \right) &=&
    \frac{g^2}{64} \left( T_a T_b + 3 T_b T_a \right)_{ji}.
\end{eqnarray} \end{subequations}

\noindent Note that if $V^a=V^b$, then the results~\eqref{uti} must be divided by $\sqrt{2}$.

\subsection{Scalar scattering}

In this section, we calculate the zeroth partial-wave of the process
\begin{equation}
	\varphi_i \, \varphi^\dagger_j \to V^a V^b.
\end{equation}

\noindent Four diagrams contribute to the corresponding amplitude:

\begin{equation}
\begin{split}
i\mathcal{M}(ij\to ab) \ = &\  \begin{tikzpicture}[baseline={([yshift=-1pt]current bounding box.center)},vertex/.style={anchor=base,
    circle,fill=black!25,minimum size=18pt,inner sep=2pt}]
\begin{feynman}
	\vertex (a1) at (-1.7,1.) {$\varphi_{i}$};
	\vertex (a2) at (0,0.8) ;
	\vertex (a3) at (1.7,1.) {$V^a_\mu$};
	\vertex (b1) at (-1.7,-1.) {$\varphi_{j}^\dagger$};
	\vertex (b2) at (0,-0.8) ;
	\vertex (b3) at (1.7,-1.) {$V^b_\nu$};
	\diagram*{ (a1) -- [scalar, momentum=\( p_1 \)](a2),
	(a2) -- [photon, momentum=\( p_3 \)](a3),
	(a2) -- [scalar, edge label=\( \varphi_k \)](b2),
	(b1) -- [scalar, momentum'=\( p_2 \)](b2),
	(b2) -- [photon, momentum'=\( p_4 \)](b3),};
\end{feynman}
\end{tikzpicture}\ \ +\ \ 
\begin{tikzpicture}[baseline={([yshift=-1pt]current bounding box.center)},vertex/.style={anchor=base,
    circle,fill=black!25,minimum size=18pt,inner sep=2pt}]
\begin{feynman}
	\vertex (a1) at (-1.7,1.) {$\varphi_{i}$};
	\vertex (a2) at (0,0.8) ;
	\vertex (a3) at (1.7,1.) {$V^b_\nu$};
	\vertex (b1) at (-1.7,-1.) {$\varphi_{j}^\dagger$};
	\vertex (b2) at (0,-0.8) ;
	\vertex (b3) at (1.7,-1.) {$V^a_\mu$};
	\diagram*{ (a1) -- [scalar, momentum=\( p_1 \)](a2),
	(a2) -- [photon, momentum=\( p_4 \)](a3),
	(a2) -- [scalar, edge label=\( \varphi_k \)](b2),
	(b1) -- [scalar, momentum'=\( p_2 \)](b2),
	(b2) -- [photon, momentum'=\( p_3 \)](b3),};
\end{feynman}
\end{tikzpicture}\ \ +\ \ \\
& \ \ +\ \ \begin{tikzpicture}[baseline={([yshift=-1pt]current bounding box.center)},vertex/.style={anchor=base,
    circle,fill=black!25,minimum size=18pt,inner sep=2pt}]
\begin{feynman}
	\vertex (a1) at (-1.7,1.1) {$\varphi_{i}$};
	\vertex (a2) at (-0.9,0) ;
	\vertex (a3) at (1.7,1.1) {$V_\mu^a$};
	\vertex (b1) at (-1.7,-1.1) {$\varphi^\dagger_{j}$};
	\vertex (b2) at (0.9,0) ;
	\vertex (b3) at (1.7,-1.1) {$V_\nu^b$};
	\diagram*{ (a1) -- [scalar, momentum=\( p_1 \)](a2),
	(b1) -- [scalar, momentum'=\( p_2 \)](a2),
	(a2) -- [photon, edge label=\( V^c \)](b2),
	(b2) -- [photon, momentum=\( p_3 \)](a3),
	(b2) -- [photon, momentum'=\( p_4 \)](b3),};
\end{feynman}
\end{tikzpicture} 
\ \ +\ \ \begin{tikzpicture}[baseline={([yshift=-1pt]current bounding box.center)},vertex/.style={anchor=base,
    circle,fill=black!25,minimum size=18pt,inner sep=2pt}]
\begin{feynman}
	\vertex (a1) at (-1.2,1.2) {$\varphi_{i}$};
	\vertex (a3) at (1.2,1.2) {$V_\mu^a$};
	\vertex (c)  at (0,0);
	\vertex (b1) at (-1.2,-1.2) {$\varphi^\dagger_{j}$};
	\vertex (b3) at (1.2,-1.2) {$V_\nu^b$};
	\diagram*{ (a1) -- [scalar, momentum=\( p_1 \)](c),
	(b1) -- [scalar, momentum=\( p_2 \)](c),
	(c) -- [photon, momentum'=\( p_3 \)](a3),
	(c) -- [photon, momentum'=\( p_4 \)](b3),};
\end{feynman}
\end{tikzpicture}.
\end{split}
\label{diags_scalar}
\end{equation}

\vspace{10pt}

\noindent We work in the high-energy limit and we therefore neglect
the masses of the scalars and gauge bosons.
In that limit,
if the two final-state gauge bosons have opposite polarizations,
then the sum of the four diagrams in Eq.~\eqref{diags_scalar} vanishes.
On the other hand,
if the two final-state gauge bosons have the same polarization,
then both the third and fourth diagrams in Eq.~\eqref{diags_fermion} vanish
and only the first two diagrams are nonzero.
The amplitudes for those two diagrams read
\begin{subequations}
\begin{align}
  \mathcal{M}_t \left( ij\to ab \right) &=
  g^2 \left( T_b T_a \right)_{ji}
  \frac{\left[ \left( 2 p_2 - p_4 \right) \cdot
      \varepsilon^*\left( p_4 \right) \right]
    \left[ \left( 2 p_1 - p_3 \right) \cdot \varepsilon^*
      \left( p_3 \right) \right]}{t},
\label{ss_tchan}\\
\mathcal{M}_u \left( ij\to ab \right) &= g^2 \left( T_a T_b \right)_{ji}
\frac{\left[ \left( 2 p_2 - p_3 \right) \cdot \varepsilon^* \left( p_3 \right)
    \right] \left[
    \left( 2 p_1 - p_4 \right) \cdot \varepsilon^* \left( p_4 \right)
    \right]}{u},
\label{ss_uchan}
\end{align}
\end{subequations}

\noindent respectively. Once again,
either $V^a$ and $V^b$ are both right-handed (RR configuration),
or they are both (LL configuration).
For each scenario, using the results of Appendix~\ref{sec:scalar_explicit_calculation} we arrive at the following total amplitudes:
\be
\mathcal{M}_{\textup{RR}}(ij\to ab)\ =\  
\mathcal{M}_{\textup{LL}}(ij\to ab) \  =\  -2 g^2  \left[  \left( T_aT_b \right)_{ji} \sin^2 \frac{\theta}{2}   +  \left( T_bT_a \right)_{ji} \cos^2 \frac{\theta}{2}  \right].
\ee

\noindent Therefore, using Eq.~\eqref{aaa000} and
\be
\int_{-1}^{+1} \sin^2{\frac{\theta}{2}}\ \mathrm{d} \cos{\theta} 
=\int_{-1}^{+1} \cos^2{\frac{\theta}{2}}\ \mathrm{d} \cos{\theta}
\ =\  1,
\ee
the corresponding zeroth
partial waves read
\begin{equation}
  \label{uty}
  a_0^{\textup{RR}} \left( ij\to ab \right)
  = a_0^{\textup{LL}} \left( ij\to ab \right) =
    - \frac{g^2}{16 \pi} \left(  T_a T_b + T_b T_a \right)_{ji}.
\end{equation}
These partial waves
must be divided by $\sqrt{2}$ when $V^a = V^b$. Furthermore, if $\textbf{\textit{n}}_\varphi$ is a real representation and $\varphi_i = \varphi_j$, then one must once again divide by $\sqrt{2}$. Notice the presence of $\pi$ in the denominator of Eq.~\eqref{uty},
while $\pi$ is absent from Eqs.~\eqref{uti}.

\section{Coupled-channel analysis}
\label{sec:sec2}

\subsection{Matrix of zeroth partial-waves}

As stated before,
the strongest constraints are achieved
by enforcing the unitarity condition~\eqref{unitarity_condition}
on \emph{the largest} (\emph{in modulus}) \emph{eigenvalue}
of a coupled-channel matrix of zeroth
partial waves.
In general, that largest eigenvalue\footnote{When we say
``largest eigenvalue'' we really mean the eigenvalue with the largest modulus.}
corresponds to an eigenvector that couples symmetrically,
\textit{i.e.}\ in group-invariant fashion,
both pairs of initial states and pairs of final states.
The concept is illustrated
by the example in Appendix~\ref{appen:a0_example2}.
We define the symmetric linear combinations of states as
\begin{align}
  \left[ \textbf{\textit{n}}_\psi \textbf{\textit{n}}_\psi
    \right]^{\textup{sym}} &\equiv \frac{1}{\sqrt{n}}\,
  \sum_{i=1}^n \psi_i\, \bar \psi_i,&
  \left[ VV \right]^{\textup{sym}}_{\textup{RR/LL}}
  &\equiv \frac{1}{\sqrt{2 \tilde{n}}}\,
  \sum_{a=1}^{\tilde{n}} \left( V^a V^a \right)_{\textup{RR/LL}}.
\label{symmetric_combo}
\end{align} 
The factor $1 \left/ \sqrt2 \right.$
in the second Eq.~\eqref{symmetric_combo}
accounts for the identical gauge bosons in the final state.
We \emph{assume}---based on our experience in particular cases,
like the one in Appendix~\ref{appen:a0_example2}---that
the process with the largest zeroth
partial wave is\footnote{If
our assumption turns out not to be true in some cases,
then the unitarity bounds that we derive will not be the strongest ones,
yet they will remain \emph{necessary} conditions for unitarity.}
$\left[ \textbf{\textit{n}}_\psi \textbf{\textit{n}}_\psi \right]^{\textup{sym}}
\to \left[ VV \right]^{\textup{sym}}_{\textup{RR/LL}}$.
In order to find the corresponding eigenvalues, we move to the basis
\begin{equation}
  \left\{ \left[ VV \right]^{\textup{sym}}_{\textup{RR}},
  \left[ VV \right]^{\textup{sym}}_{\textup{LL}},
  \left[ \textbf{\textit{n}}_\psi \textbf{\textit{n}}_\psi\right]^{\textup{sym}}
  \right \},
\end{equation}
in which the coupled-channel matrix takes the form
\begin{equation}
  a_0^\psi=\begin{pmatrix}
  \textbf{0}_{2 \times 2} & M \\
  {M}^T & \textbf{0}
  \end{pmatrix},
  \qquad
  M = \begin{pmatrix}
    a_0 \left( \left[ \textbf{\textit{n}}_\psi
      \textbf{\textit{n}}_\psi \right]^{\textup{sym}} \to
    \left[ VV \right]^{\textup{sym}}_{\textup{RR}} \right)
    \\[-2.ex]
    \\
    a_0 \left( \left[ \textbf{\textit{n}}_\psi
      \textbf{\textit{n}}_\psi \right]^{\textup{sym}} \to
    \left[ VV \right]^{\textup{sym}}_{\textup{LL}} \right)
  \end{pmatrix},
\label{eq:a0_matrix_reduced}
\end{equation}
where
\begin{equation}
  \label{uiv}
  a_0 \left( \left[ \textbf{\textit{n}}_\psi
    \textbf{\textit{n}}_\psi \right]^{\textup{sym}} \to
  \left[ VV \right]^{\textup{sym}}_{\textup{RR/LL}} \right)
  = \frac{1}{\sqrt{2 n \tilde{n}}}\, \sum_{i=1}^{n} \sum_{a=1}^{\tilde{n}}
  a_0^{\textup{RR/LL}} \left( ii\to aa \right).
\end{equation}
We derive explicit expressions for these matrix elements
by resorting to the group-theoretical relation~\cite{Georgi:1982jb}
\begin{equation}
  \sum_{a=1}^{\tilde{n}} T_a T_a = C \left( \textit{\textbf{n}} \right)
  \times \textbf{1}_{n \times n},
\end{equation}
where
$C \left( \textit{\textbf{n}} \right)$
is the quadratic Casimir invariant of the
$n$-dimensional representation
$\textit{\textbf{n}}$
and $\textbf{1}_{n \times n}$ is the $n \times n$ identity matrix.
From Eqs.~\eqref{iuprd} and~\eqref{uiv} we derive
\begin{subequations}
\begin{eqnarray}
  a_0 \left( \left[ \textbf{\textit{n}}_\psi
    \textbf{\textit{n}}_\psi \right]^{\textup{sym}} \to
  \left[ VV \right]^{\textup{sym}}_{\textup{RR}} \right)
  &=& - \frac{1}{\sqrt{2 n \tilde{n}}}\, \frac{g^2}{16}\,
  \sum_{i=1}^{n} \sum_{a=1}^{\tilde{n}} \left( T_a T_a \right)_{ii},\\
  &=& - \frac{1}{\sqrt{2 n \tilde{n}}}\, \frac{g^2}{16}\,
  \sum_{a=1}^{\tilde{n}} \textup{Tr} \left( T_a T_a \right)\\
  &=& - \frac{1}{\sqrt{2 n \tilde{n}}}\, \frac{g^2}{16}\,
  C \left( \textbf{\textit{n}}_\psi \right)
  \textup{Tr} \left( \textbf{1}_{n \times n} \right)\\
  &=& - \frac{g^2}{16}\, \sqrt{\frac{n}{2 \tilde{n}}}\,
  C \left( \textbf{\textit{n}}_\psi \right).
\end{eqnarray}
\label{eq:a0_psi_RL}
\end{subequations}
In a similar fashion, the other matrix elements are
\begin{equation}
a_0 \left( \left[ \textbf{\textit{n}}_\psi
    \textbf{\textit{n}}_\psi \right]^{\textup{sym}} \to
\left[ VV \right]^{\textup{sym}}_{\textup{LL}} \right) =
\frac{g^2}{16}\, \sqrt{\frac{n}{2 \tilde{n}}}\,
C \left( \textbf{\textit{n}}_\psi \right)
\label{eq:a0_psi_LL}
\end{equation}
and
\begin{equation}
a_0 \left( \left[ \textbf{\textit{n}}_\varphi
    \textbf{\textit{n}}_\varphi \right]^{\textup{sym}} \to
\left[ VV \right]^{\textup{sym}}_{\textup{RR}} \right)
=  a_0 \left( \left[ \textbf{\textit{n}}_\varphi
    \textbf{\textit{n}}_\varphi \right]^{\textup{sym}} \to
\left[ VV \right]^{\textup{sym}}_{\textup{LL}} \right)
= - \frac{g^2}{8 \pi}\, \sqrt{\frac{n}{2 \tilde{n}}}\,
C \left( \textbf{\textit{n}}_\varphi \right). \label{eq:a0_phi}
\end{equation}

We plug Eqs.~\eqref{eq:a0_psi_RL} and~\eqref{eq:a0_psi_LL}
into Eq.~\eqref{eq:a0_matrix_reduced}
and we find the matrix of zeroth
partial waves to be
\begin{equation}
  a_0^\psi = \frac{g^2}{16}\, \sqrt{\frac{n}{2 \tilde{n}}}\,
  C \left( \textbf{\textit{n}}_\psi \right)
  \begin{pmatrix} 0 & 0 & -1 \\ 0 & 0 & 1 \\ -1 & 1 & 0 \end{pmatrix}.
\end{equation}
Its largest eigenvalue of course is
\begin{equation}
  \left( a_0^\psi \right)^\mathrm{max} =
  \frac{g^2}{16}\, \sqrt{\frac{n}{\tilde{n}}}\,
  C \left( \textbf{\textit{n}}_\psi \right).
\label{max_a0_psi}
\end{equation}
Note that considering simultaneously
the final states $\left[ VV \right]^{\textup{sym}}_{\textup{RR}}$ and $\left[ VV \right]^{\textup{sym}}_{\textup{LL}}$ is effectively equivalent to multiplying by a factor $\sqrt{2}$
the zeroth partial wave of either.
Henceforward, we use this trick to account for both polarization
configurations
and we drop the subscripts RR and LL.

For a scalar multiplet, we have instead
\begin{equation}
  \left( a_0^\varphi \right)^\mathrm{max} =
  \frac{g^2}{8 \pi}\, \sqrt{\frac{n}{\tilde{n}}}\,
  C \left( \textbf{\textit{n}}_\varphi \right).
\label{max_a0_phi}
\end{equation}
The difference between the scalar and fermionic case
is just a factor $\pi / 2$:
\begin{equation}
  \left( a_0^\psi \right)^\mathrm{max}
  = \frac{\pi}{2} \left( a_0^\varphi \right)^\mathrm{max}.
\end{equation}
The factor $\pi$ arises from the integrations in Eqs.~\eqref{jvppp}.

\subsection{Multiple Particles and Multiple Symmetry Groups}

The above analysis has assumed the existence of only one matter (either fermionic or scalar) multiplet transforming non-trivially under only one symmetry group $\mathcal{G}$. We extend this situation to a scenario where there are $N_F$ fermionic multiplets $F_1, F_2, \ldots$ and $N_S$ scalar multiplets $S_1, S_2, \ldots$, that transform under a
single
symmetry group $\mathcal{G}$. In this scenario, the matrix of zeroth
partial waves
is defined in the basis\footnote{Here we account for both gauge-boson polarizations by adding the extra $\sqrt{2}$ to the zeroth partial wave.}:
\begin{equation}
  \left\{ \left[ VV \right]^{\textup{sym}},
  \left[ F_1F_1 \right]^{\textup{sym}},
  \left[ F_2F_2 \right]^{\textup{sym}},
  \ldots,
  \left[ S_1S_1 \right]^{\textup{sym}},
  \left[ S_2S_2 \right]^{\textup{sym}},
  \ldots \right\},
\end{equation}
The largest eigenvalue of the corresponding matrix of zeroth partial-waves
is then simply given by
\bs
\label{max_a0_multiple_A}
\ba
\left( a_0 \right)^\mathrm{max} &=&
\sqrt{ \sum_{i=1}^{N_F} \left[ \left( a_0^\psi \right)^\mathrm{max} \right]^2
  +  \sum_{i=1}^{N_S} \left[ \left( a_0^\varphi \right)^\mathrm{max} \right]^2}
\\ &=& \frac{g^2}{8\pi \sqrt{\tilde{n}}}\,
\sqrt{ \frac{\pi^2}{4}\, \sum_{i=1}^{N_F} n_{F_i}
  \left[ C \left( \textbf{\textit{n}}_{F_i} \right) \right]^2
  + \sum_{i=1}^{N_S} n_{S_i}
  \left[ C \left( \textbf{\textit{n}}_{S_i} \right) \right]^2},
\label{mjfk}
\ea
\es
where we have used the results
of Eqs.~\eqref{max_a0_psi} and~\eqref{max_a0_phi}
for $\left( a_0^\psi \right)^\mathrm{max}$
and $\left( a_0^\varphi \right)^\mathrm{max}$, respectively.
In Eq.~\eqref{mjfk},
$n_{F_i}$ is the number of fermions in the irreducible representation
$\textbf{\textit{n}}_{F_i}$ of the gauge group,
and $n_{S_i}$ is the number of fermions in the irreducible representation
$\textbf{\textit{n}}_{S_i}$ of the gauge group.

This formalism may be further extended to a scenario where
there are multiple matter (either fermionic or scalar)
multiplets $A, B, C, D, \ldots$ transforming under the direct product
of multiple symmetry groups
$\mathcal{G} \equiv \mathcal{G}_1 \times \mathcal{G}_2 \times
\mathcal{G}_3 \times \mathcal{G}_4 \times \cdots$.
The representation of matter field $A$ under $\mathcal{G}$ is written
\begin{equation}
  A \sim \left(
  \textbf{\textit{n}}_{A}^{\mathcal{G}_1} ,
  \textbf{\textit{n}}_{A}^{\mathcal{G}_2},
  \textbf{\textit{n}}_{A}^{\mathcal{G}_3},
  \textbf{\textit{n}}_{A}^{\mathcal{G}_4},
  \cdots \right).
\label{convention_matter_reps}
\end{equation}
We work in the basis
\begin{equation}
  \left \{ \left[VV\right]_{\mathcal{G}_1}^{\textup{sym}},\,
  \left[VV\right]_{\mathcal{G}_2}^{\textup{sym}},\,
  \left[VV\right]_{\mathcal{G}_3}^{\textup{sym}},\,
  \ldots,\,
  \left[AA\right]^{\textup{sym}},\,
  \left[BB\right]^{\textup{sym}},\,
  \left[CC\right]^{\textup{sym}},\,
  \ldots   \right \},
  \label{icv000}
\end{equation}
where $\left[VV\right]_{\mathcal{G}_i}^{\textup{sym}}$
is the symmetric linear combination of gauge bosons
of the group $\mathcal{G}_i$,
as defined in Eq.~\eqref{symmetric_combo}.\footnote{We insert an extra
factor $\sqrt{2}$ to account for both gauge-boson polarizations.}
If the matter multiplet $A$
transforms non-trivially under the gauge group $\mathcal{G}_i$,
then there are
\begin{equation}
  \label{neqi}
  \prod_{k \neq i} n_{A}^{\mathcal{G}_k} \equiv
  n_A^{\mathcal{G}_1} \times n_A^{\mathcal{G}_2}
\times \cdots \times n_A^{\mathcal{G}_{i-1}} \times n_A^{\mathcal{G}_{i+1}}
\times \cdots
\end{equation}
copies of that multiplet of the group $\mathcal{G}_i$
which interact with
$\left[ VV \right]_{\mathcal{G}_i}^{\textup{sym}}$.
We keep the original definition of $\left[AA\right]^{\textup{sym}}$
provided in Eq.~\eqref{symmetric_combo}, 
but multiply the corresponding zeroth partial wave 
by the square root of the factor
in Eq.~\eqref{neqi}.
In the basis~\eqref{icv000},
the full matrix of zeroth
partial waves is written as
\begin{equation}
  a_0=\begin{pmatrix}
  \textbf{0} & M^\prime \\
  {M^\prime}^T & \textbf{0}
  \end{pmatrix},
  \qquad \mbox{with} \quad
  M^{ \prime} = \begin{pmatrix}
\left(a_0^A\right)^{\mathrm{max}}_{\mathcal{G}_1} & \left(a_0^B\right)^{\mathrm{max}}_{\mathcal{G}_1} & \left(a_0^C\right)^{\mathrm{max}}_{\mathcal{G}_1} & \cdots \\ 
\left(a_0^A\right)^{\mathrm{max}}_{\mathcal{G}_2} & \left(a_0^B\right)^{\mathrm{max}}_{\mathcal{G}_2} & \left(a_0^C\right)^{\mathrm{max}}_{\mathcal{G}_2} & \cdots \\ 
\left(a_0^A\right)^{\mathrm{max}}_{\mathcal{G}_3} & \left(a_0^B\right)^{\mathrm{max}}_{\mathcal{G}_3} & \left(a_0^C\right)^{\mathrm{max}}_{\mathcal{G}_3} & \cdots \\ 
\vdots &  \vdots & \vdots &  \ddots\\ 
  \end{pmatrix}.
\label{a0_general}
\end{equation}
Each matrix element of $M^\prime$ is defined as
\begin{equation}
  \left( a_0^A \right)^{\mathrm{max}}_{\mathcal{G}_i}
  \equiv \frac{g_{\mathcal{G}_i}^2}{8 \pi \sqrt{\tilde n_{\mathcal{G}_i}}}\,
  \sqrt{\prod_{k} n_{A}^{\mathcal{G}_k} }\
    \times \left\{  
    \begin{array}{ll}
      \displaystyle{\frac{\pi}{2}\,
        C \left( \textbf{\textit{n}}_A^{\mathcal{G}_i} \right)} &
      \quad \Leftarrow A=\psi,\\
      C \left (\textbf{\textit{n}}_A^{\mathcal{G}_i} \right) &
      \quad \Leftarrow A=\varphi,
    \end{array}
    \right.
\label{general_matrix_element}
\end{equation}
where $g_{\mathcal{G}_i}$ is the gauge coupling constant of $\mathcal{G}_i$, $\tilde{n}_{\mathcal{G}_i}$ the dimension of its adjoint representation, and $C(\textbf{\textit{n}}_A^{\mathcal{G}_i})$ the quadratic Casimir invariant of the representation of the matter field $A$ under $\mathcal{G}_i$.
Notice that the product in Eq.~\eqref{general_matrix_element}
runs over \emph{all} $k$,
not just over $k \neq i$ as in Eq.~\eqref{neqi};
the extra factor $n_A^{\mathcal{G}_i}$ comes from Eqs.~\eqref{max_a0_psi} or \eqref{max_a0_phi}.

\section{Results}
\label{sec:sec3}

In the following sections, we will calculate
the largest eigenvalue of the matrix of zeroth
partial waves
for the gauge groups $SU(3) \times SU(2) \times U(1)$ (\textit{i.e.} the SM),
$SU(5)$,
$SO(10)$,
and $E_6$.
We shall impose the unitarity bound~\eqref{unitarity_condition}
to constrain the dimension of additional fermionic or scalar multiplets
in various scenarios.
We follow the group-theory conventions
of {\fontfamily{lmtt}\selectfont Mathematica}'s package
{\fontfamily{lmtt}\selectfont GroupMath}~\cite{Fonseca:2020vke}.
The Dynkin indices and the quadratic Casimir invariants
for the relevant irreducible representations of the various gauge groups
are listed in Appendix~\ref{appen:group_theory}.
A more comprehensive compilation of group-theory data
may be found in Ref.~\cite{Yamatsu:2015npn}.

\subsection{$SU(3) \times SU(2) \times U(1)$}

The SM is invariant under the gauge group
$\mathcal{G}_{\textup{SM}} \equiv SU(3)_\mathrm{colour} \times SU(2)_L \times U(1)_Y$.
Using the convention in Eq.~\eqref{convention_matter_reps},
the matter fields of the SM transform as
\bs
\label{SM_fermions}
\allowdisplaybreaks
\ba
L_{L,f} &\sim& \left( \textbf{1},\, \textbf{2},\, - \frac{1}{2} \right),
\label{SM_fermions_10} \\
Q_{L,f} &\sim& \left( \textbf{3},\, \textbf{2},\, \frac{1}{6} \right),
\label{SM_fermions_20}\\
\ell_{R,f} &\sim& \left( \textbf{1},\, \textbf{1},\, -1 \right),
\label{SM_fermions_1} \\
u_{R,f} &\sim& \left( \textbf{3},\, \textbf{1},\, \frac{2}{3} \right),
\label{SM_fermions_2}\\
d_{R,f} &\sim& \left( \textbf{3},\, \textbf{1},\, - \frac{1}{3} \right),
\label{SM_fermions_30}\\
H &\sim& \left( \textbf{1},\, \textbf{2},\, \frac{1}{2} \right).
\label{SM_fermions_3}
\ea
\es
There are three copies of each
of the multiplets~\eqref{SM_fermions_10}--\eqref{SM_fermions_30},
which we label through subscripts $f=1,2,3$.
An appropriate basis for the scattering states is
\ba
& &
\left\{
\left[ VV \right]_{SU(3)}^{\textup{sym}},\
\left[ VV \right]_{SU(2)}^{\textup{sym}},\
\left[ VV \right]_{U(1)}^{\textup{sym}},
\right. \no & &
\left[ L_{L,1} L_{L,1} \right]^\mathrm{sym},\
\left[ Q_{L,1} Q_{L,1} \right]^\mathrm{sym},\
\left[ \ell_{R,1} \ell_{R,1} \right]^\mathrm{sym},\
\left[ u_{R,1} u_{R,1} \right]^\mathrm{sym},\
\left[ d_{R,1} d_{R,1} \right]^\mathrm{sym},
\no & &
\left[ L_{L,2} L_{L,2} \right]^\mathrm{sym},\
\left[ Q_{L,2} Q_{L,2} \right]^\mathrm{sym},\
\left[ \ell_{R,2} \ell_{R,2} \right]^\mathrm{sym},\
\left[ u_{R,2} u_{R,2} \right]^\mathrm{sym},\
\left[ d_{R,2} d_{R,2} \right]^\mathrm{sym},
\no & & \left.
\left[ L_{L,3} L_{L,3} \right]^\mathrm{sym},\
\left[ Q_{L,3} Q_{L,3} \right]^\mathrm{sym},\
\left[ \ell_{R,3} \ell_{R,3} \right]^\mathrm{sym},\
\left[ u_{R,3} u_{R,3} \right]^\mathrm{sym},\
\left[ d_{R,3} d_{R,3} \right]^\mathrm{sym},\
\left[ HH \right]^\mathrm{sym}
\vphantom{\left[ VV \right]_{SU(3)}^{\textup{sym}},} \right\}.
\label{basis_vector_SM}
\ea
The corresponding matrix $M^\prime$,
in the notation of Eq.~\eqref{a0_general},
is $3 \times 16$ and may be written
in terms of a $3 \times 5$ sub-matrix $M^\prime_F$
and a $3 \times 1$ sub-matrix $M^\prime_H$ as
\be
M^\prime =
\left( M^\prime_F, \ \  M^\prime_F, \ \  M^\prime_F, \ \ M^\prime_H \right),
\ee
with
\bs
\ba
M_F^\prime &=& \frac{1}{576} 
\begin{pmatrix}
  0 &
  24 \sqrt{3}\, g_{SU(3)}^2 &
  0 &
  12 \sqrt{6}\, g_{SU(3)}^2 &
  12 \sqrt{6}\, g_{SU(3)}^2
  \\[4pt]
  9 \sqrt{6}\, g_{SU(2)}^2 &
  27 \sqrt{2}\, g_{SU(2)}^2 &
  0 &
  0 &
  0
  \\[4pt]
  9 \sqrt{2}\, g_{U(1)}^2 &
  \sqrt{6}\, g_{U(1)}^2 &
  36\, g_{U(1)}^2 &
  16 \sqrt{3}\, g_{U(1)}^2 &
  4 \sqrt{3}\, g_{U(1)}^2,
\end{pmatrix},
\\[10pt]
M^\prime_H &=&
\frac{1}{32 \pi}\begin{pmatrix}
0 \\ 
\sqrt{6}\, g_{SU(2)}^2 \\ 
\sqrt{2}\, g_{U(1)}^2
\end{pmatrix}.
\ea
\es
Each matrix element of $M^\prime_F$ and $M^\prime_H$
has been computed by using Eq.~\eqref{general_matrix_element}.

We use the gauge coupling constants measured at the Fermi scale
$\sqrt{s} \approx 10^2$\,GeV~\cite{ParticleDataGroup:2022pth}:
\be
g_{SU(3)} \approx 1.23, \quad\quad\quad g_{SU(2)} \approx 0.652, \quad\quad\quad
g_{U(1)} \approx 0.357.
\ee
We then numerically\footnote{There is an analytic formula
for the largest eigenvalue in terms of $g_{SU(3)}$,
$g_{SU(2)}$,
and $g_{U(1)}$.
However,
its size is large enough that it is not meaningful to write it down here.}
calculate the largest eigenvalue of the matrix $a_0$
of Eq.~\eqref{a0_general} to be
\begin{equation}
\left( a_0 \right)^{\mathrm{max}} \approx 0.269.
\label{a0_max_SM}
\end{equation}
We thus reassuringly conclude that the SM is perturbative,
since $a_0 = 0.269$ is well compatible with the unitarity condition
$\left| \textup{Re}\, a_0 \right| \leq 1/2$
of Eq.~\eqref{unitarity_condition}. 

If we suppose the existence of an additional fermionic multiplet
\be
\psi \sim \left( \textbf{\textit{n}}^{SU(3)}_\psi,\,
\textbf{\textit{n}}^{SU(2)}_\psi,\, Y_\psi \right),
\ee
then we must add the extra state $\left[ \psi \psi \right]^{\textup{sym}}$
to the basis vector~\eqref{basis_vector_SM}.
For each $\textbf{\textit{n}}^{SU(3)}_\psi$,
$\textbf{\textit{n}}^{SU(2)}_\psi$,
and $Y_\psi$ we use Eqs.~\eqref{a0_general} and~\eqref{general_matrix_element}
to construct the matrix $a_0$ and we then compute its largest eigenvalue.
In order to fulfill the unitarity condition
$\left| \textup{Re}\, \left(a_0\right)^{\textup{max}} \right| \leq \frac{1}{2}$,
when the dimensions of the representations of $SU(3)_\mathrm{colour}$ and $SU(2)_L$
that $\psi$ sits in are fixed,
the hypercharge $Y_\psi$ cannot be arbitrarily large.
We thus constrain the absolute value of $Y_\psi$;
the results are shown in the left panel of Table \ref{tab:GSM_Y_max_fermion}.
Obviously,
smaller $n^{SU(3)}_\psi$ and smaller $n^{SU(2)}_\psi$
allow for a larger $\left| Y_\psi \right|$.
In general,
any fermionic multiplet added to the Standard Model must satisfy
$n^{SU(3)}_\psi \leq 8$ and $n^{SU(2)}_\psi \leq 7$.

One may instead introduce an additional
(complex) scalar multiplet $\varphi$
which transforms under $\mathcal{G}_{\textup{SM}}$ as
\be
\varphi \sim \left( \textbf{\textit{n}}^{SU(3)}_\varphi,\,
\textbf{\textit{n}}^{SU(2)}_\varphi,\, Y_\varphi \right)
\ee
In a similar fashion,
$\left| Y_\varphi \right|$ may be constrained
by imposing partial-wave unitarity.
For each combination of $\textbf{\textit{n}}^{SU(3)}_\varphi$
and $\textbf{\textit{n}}^{SU(2)}_\varphi$,
the upper bound on $\left| Y_\varphi \right|$
is presented in the right panel of Table~\ref{tab:GSM_Y_max_fermion}.
We conclude that scalar multiplets are less constrained;
they must satisfy $n^{SU(3)}_\varphi \leq 10$ and $n^{SU(2)}_\varphi \leq 8$.
The first column of the right panel of Table~\ref{tab:GSM_Y_max_fermion}
corresponds to the scenario studied in Ref.~\cite{Hally:2012pu};
our constraints are,
however,
more stringent and accurate,
because we consider not only
the contributions from the extra scalar multiplet $\varphi$
but also the contributions from the SM matter fields of Eqs.~\eqref{SM_fermions}.

\begin{table}
\centering
\captionsetup{width=.8\linewidth}
\caption{Upper bounds on $\left| Y_\psi \right|$ (left panel)
  and on $\left| Y_\varphi \right|$ (right panel)
  imposed by the tree-level unitarity condition
  $\left| \textup{Re}\, a_0 \right| \leq \frac{1}{2}$,
  in an extension of the Standard Model through either
  a fermionic multiplet $\psi$ (left panel)
  or a scalar multiplet $\varphi$ (right panel).
  The symbol (---) stands for a negative,
  hence un-physical number.
}
\begin{tabular}{ c| c | c c c c |} 
\multicolumn{2}{c}{} & \multicolumn{4}{c}{$\textbf{\textit{n}}^{SU(3)}_\psi$}\\
\cline{3-6}
\multicolumn{1}{c}{} & $\left| Y_\psi \right|^{\mathrm{max}}$ &
$\textbf{1}$ & $\textbf{3}$ & $\textbf{6}$ & $\textbf{8}$ \\
\cline{2-6}
\multirow{2}{*}{\rotatebox[origin=c]{90}{$\textbf{\textit{n}}^{SU(2)}_\psi$\ \ \ \ \ \ \ \ }} & \textbf{1} & 7.91 & 5.96 & 4.41 & 4.04 \\
 &\textbf{2} & 6.65 & 4.96 & 2.65 & 2.13 \\
 &\textbf{3} & 6.00 & 4.39 & --- & --- \\
 &\textbf{4} & 5.52 & 3.87 & --- & --- \\
 &\textbf{5} & 5.05 & 2.95 & --- & --- \\
 &\textbf{6} & 4.37 & --- & --- & --- \\
 &\textbf{7} & 2.13 & --- & --- & --- \\
\cline{2-6}
\end{tabular}
\hfill
\begin{tabular}{ c| c | c c c c c |} 
\multicolumn{2}{c}{} & \multicolumn{5}{c}{$\textbf{\textit{n}}^{SU(3)}_\varphi$}\\
\cline{3-7}
\multicolumn{1}{c}{} & $|Y_\varphi|^{\mathrm{max}}$ & $\textbf{1}$ & $\textbf{3}$ & $\textbf{6}$ & $\textbf{8}$ & $\textbf{10}$ \\
\cline{2-7}
\multirow{2}{*}{\rotatebox[origin=c]{90}{$\textbf{\textit{n}}^{SU(2)}_\varphi$\ \ \ \ \ \ \ \ \ \ \ \ }} & \textbf{1} & 9.92 & 7.51 & 6.04 & 5.60 & 3.03\\
 &\textbf{2} & 8.34 & 6.29 & 4.78 & 4.40 & --- \\
 &\textbf{3} & 7.53 & 5.64 & 3.94 & 3.56 & --- \\
 &\textbf{4} & 6.97 & 5.15 & 2.86 & 2.10 & --- \\
 &\textbf{5} & 6.51 & 4.62 & --- & --- & --- \\
 &\textbf{6} & 6.03 & 3.74 & --- & --- & --- \\
 &\textbf{7} & 5.38 & --- & --- & --- & --- \\
 &\textbf{8} & 4.08 & --- & --- & --- & --- \\
\cline{2-7}
\end{tabular}
\label{tab:GSM_Y_max_fermion}
\end{table}

\subsection{Grand Unified Theory $SU(5)$}

Another interesting case study is the $SU(5)$ Grand Unified Theory (GUT).
In this model,
the three gauge coupling constants of the SM
unify at an energy scale $\sqrt{s} \approx 10^{16} \, \textup{GeV}$.
We use the following estimate for the $SU(5)$ gauge coupling constant
at the unification scale\footnote{This estimate
is conservative---some authors obtain larger values for $g_{SU(5)}$,
and then the unitarity bounds will be stronger.
The result for $g_{SU(5)}$ depends on the particle content of the GUT,
hence on the details of the Renormalization Group evolution
of the three SM gauge coupling constants
that merge into $g_{SU(5)}$.}~\cite{Giunti:1991ta, Martens:2010nm}:
\begin{equation}
g_{SU(5)} \approx 0.60.
\end{equation}

The adjoint representation of $SU(5)$ is 24-dimensional;
the $SU(5)$ GUT has
\be
\tilde{n}_{SU(5)} = 24
\ee
gauge bosons.
In addition,
there are \textit{three} generations
of two $SU(5)$ fermionic multiplets~\cite{Georgi:1974sy}:
\begin{align}
  \psi_{5,f} & \sim \overline{\textbf{5}},	&
  \psi_{10,f} & \sim \textbf{10}.
\label{su5_fermions}
\end{align}
with the generation index $f \in \left\{ 1, 2, 3 \right\}$.

A general feature of GUTs is the over-abundance of particles
in their scalar sectors.
But,
if perturbativity is to be preserved,
the largest eigenvalue of the matrix of zeroth partial-waves---given
in Eq.~\eqref{max_a0_multiple_A}---must satisfy
the unitarity condition~\eqref{unitarity_condition}.
As a result,
the number of fermions $N_F$,
the number of scalars $N_S$,
and the dimensions of their corresponding $SU(5)$ irreducible representations
(`irreps')
are constrained by
\begin{align}
  & \left| \mathrm{Re}\, \left( a_0 \right)^{\textup{max}} \right|^2
  \leq  \frac{1}{4}	&	&\Leftrightarrow 	&	&
  \sum_i^{N_S} n_{S_i}\,  C \left( \textbf{\textit{n}}_{S_i} \right)^2
  + \frac{\pi^2}{4}\, \sum_i^{N_F} n_{F_i}\,  C \left( \textbf{\textit{n}}_{F_i} \right)^2
  \leq  \frac{16 \pi^2
    \tilde{n}_{SU(5)}}{g_{SU(5)}^4} 
  \approx 29,243.	
\label{su5_unitarity_a0}
\end{align}
This implies an upper bound on the size of the scalar sector. 
Note that, for the most usual set of fermionic multiplets in \eqref{su5_fermions}, the second term
in
the left-hand side of 
Eq.~\eqref{su5_unitarity_a0}
takes the value
\be
\frac{\pi^2}{4}\, \sum_i^{N_F} n_{F_i}\,  C \left( \textbf{\textit{n}}_{F_i} \right)^2
= \frac{3\pi^2}{4}\, \left( 5\, C \left( \psi_{5} \right)^2 + 10\, C \left( \psi_{10} \right)^2 \right) \  \approx \  1,173.
\ee

We firstly
consider four minimal scenarios found in the literature~\cite{Georgi:1974sy, Georgi:1979df, Hinze:2024vrl},
all of which employ a 24-dimensional scalar multiplet
to spontaneously break $SU(5)$ into $\mathcal{G}_{\textup{SM}}$,
but differ in the set of scalars used to further break
$\mathcal{G}_{\textup{SM}}$ into the gauge group
$\mathcal{G}_{31} \equiv SU(3)_\mathrm{colour}
\times U(1)_\mathrm{electromagnetism}$
and to generate fermion masses.
For each scenario below, we use Eq.~\eqref{max_a0_multiple_A}
and the results of Appendix~\ref{appen:group_theory_su5}
to compute the largest eigenvalue of the matrix of zeroth
partial waves:
\bs
\ba
& SU(5) \xrightarrow[]{\ \ \mathbf{24}\ \ } \mathcal{G}_{\textup{SM}} \xrightarrow[]{\ \ \mathbf{5}\ \ } \mathcal{G}_{31} \ \ \Rightarrow \ \ \left( a_0 \right)^{\mathrm{max}} & \approx   0.12,\\
& SU(5) \xrightarrow[]{\ \ \mathbf{24}\ \ } \mathcal{G}_{\textup{SM}} \xrightarrow[]{\ \ \mathbf{5} \, \oplus \,  \mathbf{45}\ \ } \mathcal{G}_{31} \ \ \Rightarrow \ \ \left( a_0 \right)^{\mathrm{max}} & \approx  0.18,\\
& SU(5) \xrightarrow[]{\ \ \mathbf{24}\ \ } \mathcal{G}_{\textup{SM}} \xrightarrow[]{\mathbf{5}\,\oplus\,\mathbf{5}\,\oplus \, \mathbf{5} \,\oplus \, \mathbf{45}} \mathcal{G}_{31} \ \ \Rightarrow \ \ \left( a_0 \right)^{\mathrm{max}} & \approx  0.18,\\
& SU(5) \xrightarrow[]{\ \ \mathbf{24}\ \ } \mathcal{G}_{\textup{SM}} \xrightarrow[]{\mathbf{5} \, \oplus \, \mathbf{40} \, \oplus \, \mathbf{45}} \mathcal{G}_{31} \ \ \Rightarrow \ \ \left( a_0 \right)^{\mathrm{max}} & \approx  0.22.
\ea
\es
We find that
every scenario satisfies the unitarity condition~\eqref{unitarity_condition}.

A natural extension of these four scenarios
is the addition of either another fermionic multiplet
$\psi \sim \textbf{\textit{n}}_\psi$
or another scalar multiplet $\varphi \sim \textbf{\textit{n}}_\varphi$.
Constraints on the dimensionality of such multiplets are 
attained through Eq.~\eqref{su5_unitarity_a0}.
Regarding the former possibility,
one can immediately see
that unitarity is violated for the irrep $\textbf{70'}$
and for any fermionic multiplet with dimension $n_\psi \geq 105$. 
For scalars the bounds are less strict:
in the first three scenarios, unitarity is violated 
for the irreps $\textbf{126'}$, $\textbf{175''}$ 
and for any multiplet with dimension $n_\varphi \geq 200$;
in the fourth scenario, unitarity
is furthermore violated for the irreps $\textbf{160}$ and $\textbf{175'}$.

The five-dimensional representation of $SU(5)$
contributes very little to $\left( a_0 \right)^{\mathrm{max}}$.
We checked that,
in order to break perturbative unitarity,
one must add to the minimal $SU(5)$ GUT
close to one thousand 5-plets of scalars.

We summarize the results of this subsection in Table \ref{tab:su5_fim}.

\begin{table}
\centering
\captionsetup{width=.8\linewidth}
\caption{Summary of the $SU(5)$ extensions considered.}
\begin{tabular}{ c c c } 
\hline\\[-2.ex]
Fermion Content & Scalar Content & $\left|\textup{Re}\, \left( a_0 \right)^{\mathrm{max}}\right| \leq \frac{1}{2}$ if \\
\\[-2.ex]
\hline\hline
\\[-2.ex]
\multirow{4}{*}{$3\,(\psi_5 \oplus \psi_{10}) \oplus \textbf{\textit{n}}_\psi$}
	& $\textbf{5} \oplus \textbf{24}$   & 
	\multirow{4}{*}{$n_\psi \leq 75 \, \backslash \, \{ \textbf{70'}\}$}\\
	& $\textbf{5} \oplus \textbf{24} \oplus \textbf{45} $   & \\
	& $3(\textbf{5}) \oplus \textbf{24} \oplus \textbf{45} $   & \\
	& $\textbf{5} \oplus \textbf{24} \oplus \textbf{40} \oplus \textbf{45}$   & \\[3pt]
\hline
\\[-2.ex]
\multirow{3}{*}{$3\,(\psi_5 \oplus \psi_{10})$}	& $\textbf{5} \oplus \textbf{24} \oplus \textbf{\textit{n}}_\varphi$   & \multirow{3}{*}{$n_\varphi \leq 175 \, \backslash \, \{ \textbf{126'}, \textbf{175''}\}$}\\
	& $\textbf{5} \oplus \textbf{24} \oplus \textbf{45} \oplus \textbf{\textit{n}}_\varphi$   & \\
	& $3(\textbf{5}) \oplus \textbf{24} \oplus \textbf{45} \oplus \textbf{\textit{n}}_\varphi$   & \\[3pt]
\hline
\\[-2.ex]
$3\,(\psi_5 \oplus \psi_{10})$	& $\textbf{5} \oplus \textbf{24} \oplus \textbf{40} \oplus \textbf{45} \oplus \textbf{\textit{n}}_\varphi$   & $n_\varphi \leq 126 \vee n_\varphi=\textbf{175}$\\[3pt]
\hline
\\[-2.ex]
\multirow{3}{*}{$3\,(\psi_5 \oplus \psi_{10})$}	& $\textbf{24} \oplus N_5 (\textbf{5})$   & $N_5 \leq 953$\\
	& $\textbf{24} \oplus \textbf{45} \oplus N_5 (\textbf{5})$   & $N_5 \leq 889$\\
	&	$\textbf{24} \oplus \textbf{40} \oplus \textbf{45} \oplus N_5 (\textbf{5})$ &	$N_5 \leq 829$\\
\\[-2.ex]
\hline
\end{tabular}
\label{tab:su5_fim}
\end{table}

\subsection{Grand Unified Theory $SO(10)$}

In the $SO(10)$ GUT,
the three SM coupling constants unify
at the energy scale $\sqrt{s} \approx 10^{16} \, \textup{GeV}$
at the value~\cite{Djouadi:2022gws}
\be
g_{SO(10)} \approx 0.60.
\ee
The adjoint representation of $SO(10)$ is 45-dimensional,
thus
\be
\tilde n_{SO(10)} = 45.
\ee
Each generation of 15 SM fermions is contained
in a single 16-dimensional multiplet $\psi_{16,f} \sim \textbf{16}$
($f \in \left\{ 1,2,3 \right\}$),
which also accommodates a right-handed neutrino.
The scalar sector of the $SO(10)$ GUT may be divided in two parts:
the scalars required to break $SO(10)$ into $\mathcal{G}_{\textup{SM}}$,
and the scalars needed to generate the fermion masses
and/or to break $\mathcal{G}_{\textup{SM}}$ into $\mathcal{G}_{31}$.
There are many possible ways to break $SO(10)$ into $\mathcal{G}_{\textup{SM}}$.
The most common minimal models have one of the following symmetry-breaking
chains~\cite{PhysRevD.31.1718, Ferrari:2018rey}:
\bs
\allowdisplaybreaks
\ba
& & SO(10)  \xrightarrow[]{\mathbf{144}} \mathcal{G}_{\textup{SM}},
\label{aaaaa} \\
& & SO(10) \xrightarrow[]{\mathbf{45}}
\mathcal{G}_{421} \xrightarrow[]{\mathbf{16} /
  \mathbf{126}} \mathcal{G}_{\textup{SM}},
\\
& & SO(10) \xrightarrow[]{\mathbf{45}}
\mathcal{G}_{421} \xrightarrow[]{\mathbf{45} /
  \mathbf{210}}  \mathcal{G}_{3211}  \xrightarrow[]{\mathbf{16}
  / \mathbf{126}} \mathcal{G}_{\textup{SM}},
\\
& & SO(10) \xrightarrow[]{\mathbf{16} / \mathbf{126}}
\mathcal{G}_{5} \xrightarrow[]{\mathbf{45} / \mathbf{54}
  / \mathbf{210}} \mathcal{G}_{\textup{SM}},
\\
& & SO(10) \xrightarrow[]{\mathbf{45} / \mathbf{210}}
\mathcal{G}_{51} \xrightarrow[]{\mathbf{16} / \mathbf{126}}
\mathcal{G}_{\textup{SM}},
\\
& & SO(10) \xrightarrow[]{\mathbf{45} / \mathbf{210}} \mathcal{G}_{51}
\xrightarrow[]{\mathbf{16} / \mathbf{126}} \mathcal{G}_{5}
\xrightarrow[]{\mathbf{45} / \mathbf{54}  / \mathbf{210}}
\mathcal{G}_{\textup{SM}},
\\
& & SO(10) \xrightarrow[]{\mathbf{54} / \mathbf{210}} \mathcal{G}_{422}
\xrightarrow[]{\mathbf{16} / \mathbf{126}} \mathcal{G}_{\textup{SM}},
\\
& & SO(10) \xrightarrow[]{\mathbf{54} / \mathbf{210}}
\mathcal{G}_{422} \xrightarrow[]{\mathbf{45}} \mathcal{G}_{421}
\xrightarrow[]{\mathbf{16} / \mathbf{126}} \mathcal{G}_{\textup{SM}},
\\
& & SO(10) \xrightarrow[]{\mathbf{54} / \mathbf{210}} \mathcal{G}_{422}
\xrightarrow[]{\mathbf{45} / \mathbf{210}} \mathcal{G}_{3211}
\xrightarrow[]{\mathbf{16} / \mathbf{126}} \mathcal{G}_{\textup{SM}},
\\
& & SO(10) \xrightarrow[]{\mathbf{54} / \mathbf{210}} \mathcal{G}_{422}
\xrightarrow[]{\mathbf{45} / \mathbf{210}} \mathcal{G}_{3221}
\xrightarrow[]{\mathbf{16} / \mathbf{126}} \mathcal{G}_{\textup{SM}},
\\
& & SO(10)  \xrightarrow[]{\mathbf{45} / \mathbf{210}} \mathcal{G}_{3221}
\xrightarrow[]{\mathbf{16} / \mathbf{126}} \mathcal{G}_{\textup{SM}},
\\
& & SO(10)  \xrightarrow[]{\mathbf{45} / \mathbf{210}} \mathcal{G}_{3211}
\xrightarrow[]{\mathbf{16} / \mathbf{126}} \mathcal{G}_{\textup{SM}},
\\
& & SO(10)  \xrightarrow[]{\mathbf{45} / \mathbf{210}} \mathcal{G}_{3221}
\xrightarrow[]{\mathbf{45} / \mathbf{210}} \mathcal{G}_{3211}
\xrightarrow[]{\mathbf{16} / \mathbf{126}} \mathcal{G}_{\textup{SM}},
\ea
\label{so10_scenarios}
\es
where
\bs
\allowdisplaybreaks
\ba
\mathcal{G}_{5} &\equiv& SU(5), \\
\mathcal{G}_{51} &\equiv& SU(5) \times U(1), \\
\mathcal{G}_{421} &\equiv& SU(4) \times SU(2) \times U(1), \\
\mathcal{G}_{422} &\equiv& SU(4) \times SU(2) \times SU(2), \\
\mathcal{G}_{3211} &\equiv& SU(3) \times SU(2) \times U(1) \times U(1), \\
\mathcal{G}_{3221} & \equiv&
SU(3) \times SU(2) \times SU(2) \times U(1).
\ea
\es
Except for the direct route~\eqref{aaaaa},
all other paths employ either a $\textbf{16}$ or a $\textbf{126}$
together with a combination of $\textbf{45}$'s,
$\textbf{54}$'s,
and $\textbf{210}$'s. 

To further break $\mathcal{G}_{\textup{SM}}$ into $\mathcal{G}_{31}$,
we consider models with a $\textbf{10}$ or a $\textbf{120}$
(which generate the masses of the Standard Model quarks and leptons)
and a $\overline{\textbf{126}}$
(which furthermore generates neutrino masses via a seesaw mechanism).
Schematically:

\bs
\allowdisplaybreaks
\label{GSM_to_G31_SO10}
\ba
& & \mathcal{G}_{\textup{SM}}\xrightarrow[]{\ \mathbf{10} \, \oplus \, \overline{\mathbf{126}}\ } \mathcal{G}_{31}, \\
& & \mathcal{G}_{\textup{SM}}\xrightarrow[]{\ \mathbf{120} \, \oplus \, \overline{\mathbf{126}}\ } \mathcal{G}_{31}.
\ea
\es
We use
\be
\sum_i^{N_S} n_{S_i}\,  C \left( n_{S_i} \right)^2
+ \frac{\pi^2}{4}\, \sum_i^{N_F} n_{F_i}\,  C \left( n_{F_i} \right)^2
\ \leq \ \frac{16 \pi^2 \tilde{n}_{SO(10)}}{g_{SO(10)}^4}\  \approx \  54,831,
\label{su16_unitarity_a0}
\ee
where $N_S$ is the total number of scalars in representations $\textbf{\textit{n}}_{S_i}$
and $N_F$ is the total number of fermions in representations $\textbf{\textit{n}}_{F_i}$;
thus,
if there are three fermionic $\mathbf{16}$ representations,
\be
\sum_i^{N_S} n_{S_i}\,  C \left( n_{S_i} \right)^2
 \ \leq \ 
\frac{16 \pi^2 \tilde{n}_{SO(10)}}{g_{SO(10)}^4}
- \frac{\pi^2}{4} \times 48 \times \left( \frac{45}{8} \right)^2 \ \approx \  51,083.
\label{so10_unitarity_a0}
\ee

The condition \eqref{so10_unitarity_a0}
may
be used to evaluate the validity of perturbation theory in each scenario presented in Eqs.~\eqref{so10_scenarios} and \eqref{GSM_to_G31_SO10}. 
The results are shown in Table \ref{tab:pure_SO10},
wherein the scalar contents of the grey-shaded scenarios
violate the partial-wave unitarity condition
$\left| \textup{Re}\, \left( a_0 \right)^{\textup{max}} \right| \leq 1/2$. 
Since $\left( a_0 \right)^{\textup{max}}$ is an increasing function
of the number of scalar multiplets, 
we omit all other scenarios
for which a subset of scalars
already violates unitarity.
Thus,
we advocate a reexamination of the validity of perturbation theory
in $SO(10)$ models that have, at least, one of the following sets of scalar irreps:
\bs
\label{jvifdofp}
\ba
& & \left( \mathbf{10}\ \mbox{or higher}\right) \oplus \mathbf{210} \oplus \mathbf{210};
\\
& & \left( \mathbf{45}\ \mbox{or higher}\right) \oplus \mathbf{\overline{126}} \oplus \mathbf{210};
\\
& & \left( \mathbf{10}\ \mbox{or higher}\right)
\oplus \mathbf{120} \oplus \mathbf{126} \oplus \mathbf{\overline{126}}.
\ea
\es

\begin{table}
\centering
\captionsetup{width=.8\linewidth}
\caption{Largest zeroth
  partial waves
  of minimal $SO(10)$ GUTs.
  Every scenario assumes a fermionic sector composed of three families of $\textbf{16}$ multiplets. }
\begin{tabular}{ c c c } 
\hline\\[-2.ex]
 $\ \ \ SO(10) \to \mathcal{G}_{\textup{SM}}\ \ \ $ & $\ \ \ \mathcal{G}_{\textup{SM}} \to \mathcal{G}_{31}\ \ \ $	&	$\left( a_0 \right)^{\mathrm{max}}$ \\
\\[-2.ex]
\hline\hline
\\[-2.ex]
	$\textbf{144}$	&	$\textbf{10} \oplus \overline{\textbf{126}}$	&	0.43\\
	$\textbf{144}$	&	$\textbf{120} \oplus \overline{\textbf{126}}$	&	0.49\\

	$\textbf{16} \oplus \textbf{45}$	&	$\textbf{10} \oplus \overline{\textbf{126}}$	&	0.35\\
	$\textbf{16} \oplus \textbf{45}$	&	$\textbf{120} \oplus \overline{\textbf{126}}$	&	0.43\\
	$\textbf{16} \oplus \textbf{54}$	&	$\textbf{10} \oplus \overline{\textbf{126}}$	&	0.37\\
	$\textbf{16} \oplus \textbf{54}$	&	$\textbf{120} \oplus \overline{\textbf{126}}$	&	0.44\\
	$\textbf{16} \oplus \textbf{210}$	&	$\textbf{10} \oplus \overline{\textbf{126}}$	&	0.50\\
\rowcolor{gray!30}
	$\textbf{16} \oplus \textbf{210}$	&	$\textbf{120} \oplus \overline{\textbf{126}}$	&	0.55\\
	$\textbf{45} \oplus \textbf{126}$	&	$\textbf{10} \oplus \overline{\textbf{126}}$	&	0.46\\
\rowcolor{gray!30}
	$\textbf{45} \oplus \textbf{126}$	&	$\textbf{120} \oplus \overline{\textbf{126}}$	&	0.52\\
	$\textbf{54} \oplus \textbf{126}$	&	$\textbf{10} \oplus \overline{\textbf{126}}$	&	0.47\\
\rowcolor{gray!30}
	$\textbf{54} \oplus \textbf{126}$	&	$\textbf{120} \oplus \overline{\textbf{126}}$	&	0.53\\
\rowcolor{gray!30}
	$\textbf{126} \oplus \textbf{210}$	&	$\textbf{10} \oplus \overline{\textbf{126}}$	&	0.58\\
\rowcolor{gray!30}
	$\textbf{126} \oplus \textbf{210}$	&	$\textbf{120} \oplus \overline{\textbf{126}}$	&	0.63\\

	$\textbf{16} \oplus \textbf{45} \oplus \textbf{45}$	&	$\textbf{10} \oplus \overline{\textbf{126}}$	&	0.37\\
	$\textbf{16} \oplus \textbf{45} \oplus \textbf{45}$	&	$\textbf{120} \oplus \overline{\textbf{126}}$	&	0.44\\
	$\textbf{16} \oplus \textbf{45} \oplus \textbf{54}$	&	$\textbf{10} \oplus \overline{\textbf{126}}$	&	0.38\\
	$\textbf{16} \oplus \textbf{45} \oplus \textbf{54}$	&	$\textbf{120} \oplus \overline{\textbf{126}}$	&	0.46\\
\rowcolor{gray!30}
	$\textbf{16} \oplus \textbf{45} \oplus \textbf{210}$	&	$\textbf{10} \oplus \overline{\textbf{126}}$	&	0.51\\
\rowcolor{gray!30}
	$\textbf{16} \oplus \textbf{45} \oplus \textbf{210}$	&	$\textbf{120} \oplus \overline{\textbf{126}}$	&	0.57\\
	$\textbf{16} \oplus \textbf{54} \oplus \textbf{54}$	&	$\textbf{10} \oplus \overline{\textbf{126}}$	&	0.40\\
	$\textbf{16} \oplus \textbf{54} \oplus \textbf{54}$	&	$\textbf{120} \oplus \overline{\textbf{126}}$	&	0.47\\
\rowcolor{gray!30}
	$\textbf{16} \oplus \textbf{54} \oplus \textbf{210}$	&	$\textbf{10} \oplus \overline{\textbf{126}}$	&	0.52\\
	$\textbf{45} \oplus \textbf{45} \oplus \textbf{126}$	&	$\textbf{10} \oplus \overline{\textbf{126}}$	&	0.47\\
\\[-2.ex]
\hline
\end{tabular}
\label{tab:pure_SO10}
\end{table}

One may want to expand the scalar sector of the remaining $SO(10)$ models
by incorporating low-dimensional irreps.
Specifically,
we consider adding either $N_{10}$ copies
of ten-dimensional scalar multiplets $\textbf{10}$,
or (but not simultaneously) $N_{16}$ copies
of sixteen-dimensional scalar multiplets $\textbf{16}$.
According to Eq.~\eqref{so10_unitarity_a0},
$N_{10}$ and $N_{16}$ must be bounded from above.
In Table~\ref{tab:N10_N16_SO10}
we present upper bounds on $N_{10}$ and $N_{16}$
for some $SO(10)$ models
that do not violate tree-level partial-wave unitarity,
yet have rather low
(smaller than 10)
maximal $N_{10}$ and/or maximal $N_{16}$.

\begin{table}
\centering
\captionsetup{width=.8\linewidth}
\caption{Upper bounds on $N_{10}$ and $N_{16}$ for some allowed $SO(10)$ models.
We always assume the fermionic content to be three $\mathbf{16}$ of $SO(10)$.}
\begin{tabular}{ccc}
\hline\\[-2.ex]
$SO(10) \to \mathcal{G}_{\textup{SM}} \to \mathcal{G}_{31}$ &
$N_{10}^{\mathrm{max}}$ &
$N_{16}^{\mathrm{max}}$ \\
\\[-2.ex]
\hline\hline
\\[-13pt]
$\mathbf{120} \oplus \mathbf{126} \oplus \mathbf{144}$ & 9 & 0
\\
$\mathbf{120} \oplus \mathbf{126} \oplus \mathbf{144}$ & 0 & 3
\\\hline
$\mathbf{10} \oplus \mathbf{126} \oplus \mathbf{210}$ & 2 & 0
\\
$\mathbf{10} \oplus \mathbf{126} \oplus \mathbf{210}$ & 0 & 1
\\\hline
$\mathbf{10} \oplus \mathbf{54} \oplus \mathbf{54} \oplus \mathbf{126}
\oplus \mathbf{126}$ & 3 & 0
\\
$\mathbf{10} \oplus \mathbf{54} \oplus \mathbf{54} \oplus \mathbf{126}
\oplus \mathbf{126}$ & 0 & 1
\\
\hline
\end{tabular}
\label{tab:N10_N16_SO10}
\end{table}

\subsection{Grand Unified Theory $E_6$}

We now study the $E_6$ Grand Unified Theory.
At the GUT energy scale $\sqrt{s} \approx 10^{16} \, \textup{GeV}$,
we adopt the conservative estimate
for the unified coupling constant~\cite{Dash:2020jlc}
\be
g_{E_6} \approx 0.55.
\ee
The adjoint representation of $E_6$ is 78-dimensional,
thus
\be
\tilde n_{E_6} = 78.
\ee
The fermionic sector is composed of three generations
of a 27-dimensional multiplet $\psi_{27}^f \sim \textbf{27}$,
$f \in \left\{ 1, 2, 3 \right\}$.
Each $\mathbf{27}$ accommodates the 15 SM fermions
of Eqs.~\eqref{SM_fermions_10}--\eqref{SM_fermions_30}
along with 12 new fermions,
out of which one is a right-handed neutrino~\cite{Gursey:1975ki}.

The tensor product of two $\textbf{27}$'s decomposes as
\begin{equation}
  \textbf{27} \otimes \textbf{27} =
  \textbf{27} \oplus \textbf{351} \oplus \textbf{351'}
\end{equation}
under $E_6$.
Therefore,
fermion masses can be generated through a Higgs-like mechanism
by introducing any linear combination of $\textbf{27}$,
$\textbf{351}$,
and $\textbf{351'}$ scalar multiplets.
However,
we expect the masses of the 12 fermions beyond the SM
to be orders of magnitude larger than the masses of the SM  fermions.
This splitting of mass scales can only be achieved by including
at least one scalar multiplet
$\textbf{351'}$~\cite{Ramond:1979py, Schwichtenberg:2017xhv, Haber:1986gz}.
With this in mind,
before specifying the scalar multiplets
needed to break $E_6$ into $\mathcal{G}_{\textup{SM}}$,
we compute the largest eigenvalue of the matrix of zeroth partial-waves
for a model composed of three $\textbf{27}$ fermionic multiplets
and one $\textbf{351'}$ scalar multiplet.
We define the matrix of zeroth partial-waves in the basis:
\be
\left\{ \left[ VV \right]^{\textup{sym}}_{E_6},\,
\left[ \psi_{27}^1\, \psi_{27}^1 \right]_{\textup{sym}},\,
\left[ \psi_{27}^2\, \psi_{27}^2 \right]_{\textup{sym}},\,
\left[ \psi_{27}^3\, \psi_{27}^3 \right]_{\textup{sym}},\,
\left[ \textbf{351'}\, \textbf{351'} \right]_{\textup{sym}} \right\}.
\ee
By using Eq.~\eqref{max_a0_multiple_A}
we immediately find that the largest eigenvalue
violates the unitarity condition of Eq.~\eqref{unitarity_condition}:
\be
\left( a_0 \right)^{\mathrm{max}}  \ =\ 
\frac{g_{E_6}^2}{8\pi \sqrt{78}} \sqrt{ \frac{3\pi^2}{4}
  \left[\, \sqrt{27}\,C \left( \textbf{27} \right) \right]^2
  + \left[ \sqrt{351}\, C \left( \textbf{351'} \right) \right]^2} 
  \ \approx\  0.51.
\ee
The situation can only worsen with the introduction
of further scalar or fermionic multiplets.
To give one out
of many examples found in the literature,
Ref.~\cite{Joglekar:2016yap} considers a model
with three $\textbf{27}$ fermionic multiplets
and a scalar sector composed of
$\textbf{27} \oplus \textbf{78} \oplus \textbf{351'}$.
By running the three SM couplings up to the GUT energy scale,
where they unify,
Ref.~\cite{Joglekar:2016yap} obtains $g_{E_6} \approx 0.70$.
The corresponding largest eigenvalue in this scenario equals
\be
\left( a_0 \right)^{\mathrm{max}} \approx 0.86,
\ee
which obviously violates
the unitarity condition~\eqref{unitarity_condition}.
For this reason,
we raise questions about the viability
of \emph{any} perturbative Grande Unified Theory with $E_6$ gauge symmetry,
capable of generating the correct mass hierarchy in the fermion sector.

\section{Conclusion}
\label{sec:conclusion}
We have computed the largest (in modulus) eigenvalue
of the matrix of zeroth partial waves
in the scattering
of a fermion--antifermion pair
and of a scalar--scalar pair
into two gauge bosons.
We have made the educated guess that the largest eigenvalue is always obtained
when the fermion--antifermion,
scalar--scalar,
and gauge boson--gauge boson pairs
are in symmetric combinations under the gauge group;
if this assumption fails in some cases---which we do not expect
to happen---then the bounds that we have obtained will still be valid,
even if they are not the strongest possible ones.
We have found that no problem of violation of the unitarity requirement
occurs in most extensions of the SM and/or of the $SU(5)$ Grand Unified Theory.
On the other hand,
many schemes of $SO(10)$ symmetry breaking display unitarity violation,
and this is also true of \emph{all} the $E_6$-based GUTs.
In those GUTs,
perturbative calculations seem to be unwarranted.

\vspace*{5mm}

\paragraph{Acknowledgements:}
We were supported by the Portuguese Foundation for Science and Technology
through projects UIDB/00777/2020 and UIDP/00777/2020.
The work of L.L.\ was furthermore supported by projects CERN/FIS-PAR/0002/2021
and CERN/FIS-PAR/0019/2021.

\appendix

\section{Explicit calculation of the scattering amplitudes}
\label{sec:appA}

As we will be considering scattering amplitudes of processes of the type $A_i(p_1) A^\dagger_j(p_2) \to V^a(p_3) V^b(p_4)$ (with $A=\psi, \varphi$), it is important to define the four-momenta of each particle. In the $x-z$ plane these can be written as:
\bs
\begin{align}
	p_1^\mu &
	\approx \frac{\sqrt{s}}{2}(1, \sin{\theta}, 0, \cos{\theta}),\\
	p_2^\mu &
	\approx \frac{\sqrt{s}}{2}(1, -\sin{\theta}, 0, -\cos{\theta}),\\
	p_3^\mu &
	\approx \frac{\sqrt{s}}{2}(1, 0, 0, 1),\\
	p_4^\mu &
	\approx \frac{\sqrt{s}}{2}(1, 0, 0, -1),
\end{align}
\label{four_momenta}
\es
where the center-of-mass frame of reference was adopted and the high-energy limit ($\sqrt{s} \gg m$) was taken. In this limit the corresponding chiral spinors associated with $\psi_i$ and $\overline{\psi}_j$ (\textit{i.e.} $u_i$, $\overline{v}_j$) and gauge boson transverse polarization vectors ($\varepsilon^\mu_{L/R}$) can be written in the helicity basis as \cite{Thomson:2013zua,Romao:2016ien}:
\bs
\allowdisplaybreaks
\begin{align}
u_i(p_1)&\approx \frac{s^{1/4}}{\sqrt{2}}
\begin{pmatrix}
-\sin{\frac{\theta}{2}}\\ 
\cos{\frac{\theta}{2}}\\ 
\sin{\frac{\theta}{2}}\\ 
-\cos{\frac{\theta}{2}}
\end{pmatrix}, &
\overline{v}_j(p_2)&\approx \frac{s^{1/4}}{\sqrt{2}}
\begin{pmatrix}
\cos{\frac{\theta}{2}}\\ 
\sin{\frac{\theta}{2}}\\ 
\cos{\frac{\theta}{2}}\\ 
\sin{\frac{\theta}{2}}
\end{pmatrix},  \\ 
\varepsilon^\mu_L(p_3) & =\frac{1}{\sqrt2}(0, 1, - i,0),	&
\varepsilon^\mu_R(p_3) & =\frac{1}{\sqrt2}(0, 1,  i,0), \\
\varepsilon^\mu_L(p_4) & =\frac{1}{\sqrt2}(0, -1,  i,0), &
\varepsilon^\mu_R(p_4) & =\frac{1}{\sqrt2}(0, -1, - i,0).
\end{align}
\label{spinors_helicities}
\es

\subsection{Fermions}
\label{sec:fermion_explicit_calculation}

To simplify Eqs.~\eqref{ff_tchan} and \eqref{ff_uchan}, we first contract the four vectors of Eqs.~\eqref{spinors_helicities} with the Dirac gamma matrices. Working in the Pauli-Dirac representation, one obtains:
\bs
\allowdisplaybreaks
\begin{align}
\slashed{p}_1 &=\frac{\sqrt{s}}{2} \begin{pmatrix}
1 & 0 & -c_\theta & -s_\theta\\ 
0 & 1 & -s_\theta & c_\theta\\ 
c_\theta & s_\theta & -1 & 0\\ 
s_\theta & -c_\theta & 0 & -1
\end{pmatrix},	&
\slashed{p}_2 &=\frac{\sqrt{s}}{2} \begin{pmatrix}
1 & 0 & c_\theta & s_\theta\\ 
0 & 1 & s_\theta & -c_\theta\\ 
-c_\theta & -s_\theta & -1 & 0\\ 
-s_\theta & c_\theta & 0 & -1
\end{pmatrix},
\end{align}

\begin{align}
\slashed{p}_3 &=\frac{\sqrt{s}}{2} \begin{pmatrix}
1 & 0 & -1 & 0\\ 
0 & 1 & 0 & 1\\ 
1 & 0 & -1 & 0\\ 
0 & -1 & 0 & -1
\end{pmatrix},	&	
\slashed{p}_4 &=\frac{\sqrt{s}}{2} \begin{pmatrix}
1 & 0 & 1 & 0\\ 
0 & 1 & 0 & -1\\ 
-1 & 0 & -1 & 0\\ 
0 & 1 & 0 & -1
\end{pmatrix},	
\end{align}

\begin{align}
\slashed{\varepsilon}_{R}^*(p_3) &= \sqrt2\begin{pmatrix}
0 & 0 & 0 & 0\\ 
0 & 0 & -1 & 0\\ 
0 & 0 & 0 & 0\\ 
1 & 0 & 0 & 0
\end{pmatrix},	&
\slashed{\varepsilon}_{L}^*(p_3) &= \sqrt2\begin{pmatrix}
0 & 0 & 0 & -1\\ 
0 & 0 & 0 & 0\\ 
0 & 1 & 0 & 0\\ 
0 & 0 & 0 & 0
\end{pmatrix},	
\end{align}

\begin{align}
\slashed{\varepsilon}_{R}^*(p_4) &= \sqrt2\begin{pmatrix}
0 & 0 & 0 & 1\\ 
0 & 0 & 0 & 0\\ 
0 & -1 & 0 & 0\\ 
0 & 0 & 0 & 0
\end{pmatrix},	&	
\slashed{\varepsilon}_{L}^*(p_4) &= \sqrt2\begin{pmatrix}
0 & 0 & 0 & 0\\ 
0 & 0 & 1 & 0\\ 
0 & 0 & 0 & 0\\ 
-1 & 0 & 0 & 0
\end{pmatrix},
\end{align}
\es

\noindent where $c_\theta$ and $s_\theta$ are the cosine and sine of the scattering angle, respectively. The numerator of Eqs.~\eqref{ff_tchan} and \eqref{ff_uchan} can be explicitly written for the RR and LL polarization configurations as:
\bs
\allowdisplaybreaks
\begin{align}
\left[ \overline{v}_j \cdot \slashed{\varepsilon}_R^*(p_4) \cdot (\slashed{p}_1-\slashed{p}_3) \cdot \slashed{\varepsilon}_R^*(p_3) \cdot u_i \right]&=-s \sin{\theta} \sin^2{\frac{\theta}{2}},\\
\left[ \overline{v}_j \cdot \slashed{\varepsilon}_R^*(p_3) \cdot (\slashed{p}_1-\slashed{p}_4) \cdot \slashed{\varepsilon}_R^*(p_4) \cdot u_i  \right]&=-s \sin{\theta} \sin^2{\frac{\theta}{2}},\\
\left[ \overline{v}_j \cdot \slashed{\varepsilon}_L^*(p_4) \cdot (\slashed{p}_1-\slashed{p}_3) \cdot \slashed{\varepsilon}_L^*(p_3) \cdot u_i \right]&=s \sin{\theta} \cos^2{\frac{\theta}{2}},\\
\left[ \overline{v}_j \cdot \slashed{\varepsilon}_L^*(p_3) \cdot (\slashed{p}_1-\slashed{p}_4) \cdot \slashed{\varepsilon}_L^*(p_4) \cdot u_i  \right]&=s \sin{\theta} \cos^2{\frac{\theta}{2}},	
\end{align}
\label{fermion_currents}
\es

\noindent while, in the high-energy limit, the Mandelstam variables $t$ and $u$ take the form:
\begin{align}
t \equiv (p_1-p_3)^2 &\approx -s \sin^2{\frac{\theta}{2}}	,	&	u \equiv (p_1-p_4)^2 &\approx -s \cos^2{\frac{\theta}{2}}.
\label{mandelstam_high_energy}
\end{align}

Summing Eqs.~\eqref{ff_tchan} and \eqref{ff_uchan} and using Eqs.~\eqref{fermion_currents} and \eqref{mandelstam_high_energy} we arrive at the following formula for the total amplitude of the RR and LL processes:
\bs
\ba
\mathcal{M}_{\textup{RR}}(ij\to ab) &=& -g^2  \sin \theta \left [  \left (T_a T_b  \right )_{ji}   \tan^2 \frac{\theta}{2}  +  \left (T_b T_a  \right )_{ji} \right ],\\
\mathcal{M}_{\textup{LL}}(ij\to ab) &=& g^2  \sin \theta \left [  \left (T_a T_b  \right )_{ji}    +  \left (T_b T_a  \right )_{ji}  \cot^2 \frac{\theta}{2} \right ].
\ea
\es

\noindent where the group generators should be evaluated at the appropriate initial state $\psi_i \overline{\psi}_j$, as the notation implies.

\subsection{Scalars}
\label{sec:scalar_explicit_calculation}

The first step in simplifying Eqs.~\eqref{ss_tchan} and \eqref{ss_uchan} is to calculate the momenta four-vectors coupling to the polarization vectors explicitly:
\bs
\allowdisplaybreaks
\begin{align}
	(2 p_2 - p_4)^\mu &= \frac{\sqrt{s}}{2}(1, -2 \sin \theta, 0, 1-2 \cos \theta),\\
	(2 p_1 - p_3)^\mu &= \frac{\sqrt{s}}{2}(1, 2 \sin \theta, 0, -1+2 \cos \theta),\\
	(2 p_2 - p_3)^\mu &= \frac{\sqrt{s}}{2}(1, -2 \sin \theta, 0, -1-2 \cos \theta),\\
	(2 p_1 - p_4)^\mu &= \frac{\sqrt{s}}{2}(1, 2 \sin \theta, 0, 1+2 \cos \theta).
\end{align}
\es

\noindent To perform the contractions with the polarization vectors, we need to consider the RR and LL scenarios separately:
\bs
\allowdisplaybreaks
\label{ss_simplifications}
\begin{align}
	(2 p_2 - p_4)\cdot \varepsilon_R^*(p_4) &= -\sqrt{\frac{s}{2}} \sin \theta,	&	
	(2 p_1 - p_3)\cdot \varepsilon_R^*(p_3) &= -\sqrt{\frac{s}{2}} \sin \theta,		\label{ss_simplifications_1}\\
	(2 p_2 - p_4)\cdot \varepsilon_L^*(p_4) &= -\sqrt{\frac{s}{2}} \sin \theta,	&	
	(2 p_1 - p_3)\cdot \varepsilon_L^*(p_3) &= -\sqrt{\frac{s}{2}} \sin \theta,\\
	(2 p_2 - p_3)\cdot \varepsilon_R^*(p_3) &= \sqrt{\frac{s}{2}} \sin \theta,	&	
	(2 p_1 - p_4)\cdot \varepsilon_R^*(p_4) &= \sqrt{\frac{s}{2}} \sin \theta,\\
	(2 p_2 - p_3)\cdot \varepsilon_L^*(p_3) &= \sqrt{\frac{s}{2}} \sin \theta,	&	
	(2 p_1 - p_4)\cdot \varepsilon_L^*(p_4) &= \sqrt{\frac{s}{2}} \sin \theta.		\label{ss_simplifications_4}
\end{align}
\es

Summing Eqs.~\eqref{ss_tchan} and \eqref{ss_uchan} and using Eqs. \eqref{mandelstam_high_energy} and \eqref{ss_simplifications} we arrive at the following formula for the total scattering amplitude of the RR and LL processes:
\bs
\begin{align}
\mathcal{M}_{\textup{RR}}(ij\to ab) &= -2 g^2  \left[  \left( T_aT_b \right)_{ji} \sin^2 \frac{\theta}{2}   +  \left( T_bT_a \right)_{ji} \cos^2 \frac{\theta}{2}  \right],\\
\mathcal{M}_{\textup{LL}}(ij\to ab) &= -2 g^2  \left[  \left( T_aT_b \right)_{ji} \sin^2 \frac{\theta}{2}   +  \left( T_bT_a \right)_{ji} \cos^2 \frac{\theta}{2}  \right].
\end{align}
\es

\section{Instructive example of the largest eigenvalue of $a_0$}
\label{appen:a0_example2}

We consider a fermionic $\mathbf{2}_\psi$
transforming in the doublet representation of $SU(2)$.
One then has $T_a = \tau_a / 2$,
where the $\tau_a$ are the Pauli matrices.
Hence,
\bs
\allowdisplaybreaks
\ba
4 T_1 T_1 = 4 T_2 T_2 = 4 T_3 T_3
&=& \left( \begin{array}{cc} 1 & 0 \\ 0 & 1 \end{array} \right),
\\
3 T_1 T_2 + T_2 T_1 = - \left( 3 T_2 T_1 + T_1 T_2 \right)
&=& \frac{1}{2} \left( \begin{array}{cc} i & 0 \\ 0 & -i \end{array} \right),
\\
3 T_1 T_3 + T_3 T_1 = - \left( 3 T_3 T_1 + T_1 T_3 \right)
&=& \frac{1}{2} \left( \begin{array}{cc} 0 & -1 \\ 1 & 0 \end{array} \right),
\\
3 T_2 T_3 + T_3 T_2 = - \left( 3 T_3 T_2 + T_2 T_3 \right)
&=& \frac{1}{2} \left( \begin{array}{cc} 0 & i \\ i & 0 \end{array} \right).
\ea
\es

We define the matrix of zeroth partial-waves in the basis
\be
\left\{ \left[ VV \right]_{\textup{RR}},\,
\left[ VV \right]_{\textup{LL}},\,
\left[ \mathbf{2}_\psi \mathbf{2}_\psi \right] \right\},
\ee
where
\bs
\ba
\left[ VV \right]_{\textup{RR}} &=&
\frac{\left(V^1 V^1\right)_{\textup{RR}}}{\sqrt{2}},\,
\left(V^1 V^2\right)_{\textup{RR}},\,
\left(V^1 V^3\right)_{\textup{RR}},\,
\frac{\left(V^2 V^2\right)_{\textup{RR}}}{\sqrt{2}},\,
\left(V^2 V^3\right)_{\textup{RR}},\,
\frac{\left(V^3 V^3\right)_{\textup{RR}}}{\sqrt{2}}, \quad\quad
\\[10pt]
\left[ VV \right]_{\textup{LL}} &=&
\frac{\left(V^1 V^1\right)_{\textup{LL}}}{\sqrt{2}},\,
\left(V^1 V^2\right)_{\textup{LL}},\,
\left(V^1 V^3\right)_{\textup{LL}},\,
\frac{\left(V^2 V^2\right)_{\textup{LL}}}{\sqrt{2}},\,
\left(V^2 V^3\right)_{\textup{LL}},\,
\frac{\left(V^3 V^3\right)_{\textup{LL}}}{\sqrt{2}},\quad\quad
\\[10pt]
\left[ \mathbf{2}_\psi \mathbf{2}_\psi \right] &=&
\psi_1 \overline{\psi}_1,\,
\psi_2 \overline{\psi}_2,\,
\psi_1 \overline{\psi}_2,\,
\psi_2 \overline{\psi}_1.
\ea
\es
The coupled-channel matrix of zeroth partial-waves then reads
\be
\label{ldkldk}
a_0^\psi = \begin{pmatrix}
\textbf{0}_{12 \times 12} & M \\ M^T & \textbf{0}_{4 \times 4}
\end{pmatrix},
\ee
where $M$ is the following $12 \times 4$ matrix:
\begin{align}
M &= \frac{g^2}{128}\begin{pmatrix}
 - \sqrt{2} & - \sqrt{2} & 0 & 0 \\[-4pt]
 -i & i & 0 & 0 \\[-4pt]
 0 & 0 & -1 & 1 \\[-4pt]
 - \sqrt{2} & - \sqrt{2} & 0 & 0 \\[-4pt]
 0 & 0 & -i & -i \\[-4pt]
 - \sqrt{2} & - \sqrt{2} & 0 & 0 \\[-4pt]
 \sqrt{2} & \sqrt{2} & 0 & 0 \\[-4pt]
 -i & i & 0 & 0 \\[-4pt]
 0 & 0 & -1 & 1 \\[-4pt]
 \sqrt{2} & \sqrt{2} & 0 & 0 \\[-4pt]
 0 & 0 & -i & -i \\[-4pt]
 \sqrt{2} & \sqrt{2} & 0 & 0 \\[-4pt]
\end{pmatrix}.
\label{largest_eigen_4_su2}
\end{align}

The $16 \times 16$ matrix $a_0^\psi$ of Eq.~\eqref{ldkldk} has eigenvalues
\be
\pm \frac{g^2 \sqrt{3}}{32\sqrt{2}},\
\pm \frac{g^2}{64},\ 
\pm \frac{ig^2}{64},\
0.
\ee
The eigenvalue with the largest modulus is
\be
\left( a_0^\psi \right)^{\mathrm{max}} = \frac{ g^2 \sqrt{3}}{32\sqrt{2}},
\ee
and the corresponding (transposed) eigenvector is
\be
\left(
-1,\, 0,\, 0,\, 
-1,\, 0,\,
-1,\,
1,\, 0,\, 0,\,
1,\, 0,\,
1,\,
\sqrt{3},\,
\sqrt{3},\,
0,\, 0 \right)^T.
\ee
This means that the largest zeroth partial-waves arise from the processes
\bs
\ba
\left[ \left( V^1 V^1 \right)_{\textup{RR}}  +  \left( V^2 V^2 \right)_{\textup{RR}}
  + \left( V^3 V^3 \right)_{\textup{RR}} \right]
&\to & \left( \psi_1 \overline{\psi}_1  +  \psi_2 \overline{\psi}_2 \right),
\\
\left[ \left( V^1 V^1 \right)_{\textup{LL}}  +  \left( V^2 V^2 \right)_{\textup{LL}}
  + \left( V^3 V^3 \right)_{\textup{LL}} \right]
&\to & \left( \psi_1 \overline{\psi}_1  +  \psi_2 \overline{\psi}_2 \right).
\ea
\es
We have found this feature to be a general property of any representation of any symmetry group.

\section{Group Representations and Casimir Invariants}
\label{appen:group_theory}

\subsection{$U(1)$ Gauge Group}

The dimension of the $U(1)$ adjoint representation is $\tilde{n}_{U(1)} = 1$. All $U(1)$ representations are one-dimensional, and given by a charge $Y$. Table \ref{tab:u1_specs} shows these properties alongside the corresponding quadratic Casimir invariant ($C(Y)$). 

\begin{figure}
  \begin{minipage}[b]{.45\textwidth}
	\centering
	\begin{tabular}{ c c c } 
	\hline
	Charge   & Dimension & Casimir ($C(Y)$)  \\
	\hline\hline
	\\[-2.ex]
	$Y \in \mathbb{R}$  & 1 & $Y^2$ \\
	\\[-2.ex]
	\hline
	\end{tabular}
	\captionsetup{width=0.9\linewidth}
	\captionof{table}{Irreducible representations and their Casimir invariants of $U(1)$.}	
	\label{tab:u1_specs}
  \end{minipage}\hfill
  \begin{minipage}[b]{.45\textwidth}
\centering
	\begin{tabular}{ c c c c } 
	\hline
	irrep  & Dynkin & Dimension & Casimir  \\
	($\textbf{\textit{n}}$) & Index & ($n$) & ($C(\textbf{\textit{n}})$) \\
	\hline\hline
	\\[-2.ex]
	$\textbf{\textit{n}} \in \mathbb{N}$ & $\{ n-1 \}$ & $n$ & $\frac{1}{4} (n^2-1)$ \\
	\\[-2.ex]
	\hline
	\end{tabular}
	\captionsetup{width=0.9\linewidth}
	\captionof{table}{Irreducible representations and their Casimir invariants of $SU(2)$.}	
	\label{tab:su2_specs}	
  \end{minipage}\hfill
\end{figure}

\subsection{$SU(2)$ Gauge Group}

The dimension of the $SU(2)$ adjoint representation is $\tilde{n}_{SU(2)} = 3$. Table \ref{tab:su2_specs} shows the $SU(2)$ irreducible representations ($\textbf{\textit{n}}$), alongside their Dynkin indices, dimensions ($n$), and quadratic Casimir invariants ($C(\textbf{\textit{n}})$).

\subsection{$SU(3)$ Gauge Group}

\begin{figure}[h!]
  \begin{minipage}[b]{.4\textwidth}
	\centering
	\begin{tabular}{ c c c c } 
	\hline
	irrep  & Dynkin & Dimension & Casimir  \\
	($\textbf{\textit{n}}$) & Index & ($n$) & ($C(\textbf{\textit{n}})$) \\
	\hline\hline
\\[-2.ex]
\textbf{1} & $\{0,0\}$ & 1 & 0 \\
 \textbf{3} & $\{1,0\}$ & 3 & 4/3 \\
 \textbf{6} & $\{0,2\}$ & 6 & 10/3 \\
 \textbf{8} & $\{1,1\}$ & 8 & 3 \\
 \textbf{10} & $\{3,0\}$ & 10 & 6 \\
 \textbf{15} & $\{2,1\}$ & 15 & 16/3 \\
 \textbf{15'} & $\{4,0\}$ & 15 & 28/3 \\
 \textbf{21} & $\{0,5\}$ & 21 & 40/3 \\
 \textbf{24} & $\{1,3\}$ & 24 & 25/3 \\
\\[-2.ex]
	\hline
	\end{tabular}
	\vspace{10pt}
	\captionsetup{width=\linewidth}
	\captionof{table}{Lowest-dimensional $SU(3)$ irreducible representations and their Casimir invariants.}	
	\label{tab:su3_specs}
  \end{minipage}\hfill
  \begin{minipage}[b]{.5\textwidth}
    \centering
    \includegraphics[width=\linewidth]{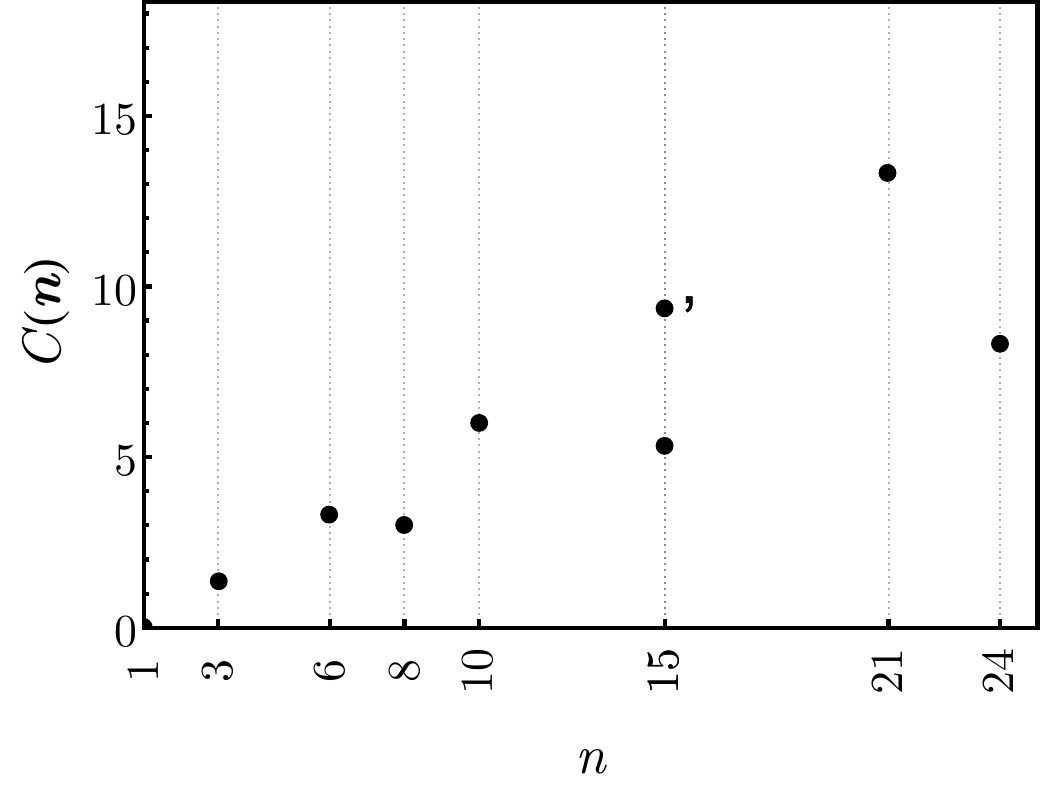}
    \captionof{figure}{$SU(3)$ Casimir invariant as a function of the dimension of the irreducible representation.}
    \label{fig:cn_su3}
  \end{minipage}\hfill
\end{figure}

The dimension of the adjoint representation is $\tilde{n}_{SU(3)} = 8$. 
Table \ref{tab:su3_specs} shows the lowest-dimensional $SU(3)$ irreducible representations ($\textbf{\textit{n}}$), alongside their Dynkin indices, dimensions ($n$), and quadratic Casimir invariants ($C(\textbf{\textit{n}})$). For a more intuitive understanding, in Figure \ref{fig:cn_su3} we plot the quadratic Casimir invariant as a function of the dimension of the irreducible representations found in Table \ref{tab:su3_specs}. The primed marker $\bullet\textbf{'}$ represents a primed representation $\textbf{\textit{n}'}$.

\subsection{$SU(5)$ Gauge Group}
\label{appen:group_theory_su5}

\begin{figure}[h!]
  \begin{minipage}[b]{.45\textwidth}
	\centering
	\begin{tabular}{ c c c c } 
	\hline
	irrep  & Dynkin & Dimension & Casimir  \\
	($\textbf{\textit{n}}$) & Index & ($n$) & ($C(\textbf{\textit{n}})$) \\
	\hline\hline
\\[-2.ex]
 \textbf{1} & $\{0,0,0,0\}$ & 1 & 0 \\
 \textbf{5} & $\{1,0,0,0\}$ & 5 & 12/5 \\
 \textbf{10} & $\{0,1,0,0\}$ & 10 & 18/5 \\
 \textbf{15} & $\{2,0,0,0\}$ & 15 & 28/5 \\
 \textbf{24} & $\{1,0,0,1\}$ & 24 & 5 \\
 \textbf{35} & $\{0,0,0,3\}$ & 35 & 48/5 \\
 \textbf{40} & $\{0,0,1,1\}$ & 40 & 33/5 \\
 \textbf{45} & $\{0,1,0,1\}$ & 45 & 32/5 \\
 \textbf{50} & $\{0,0,2,0\}$ & 50 & 42/5 \\
 \textbf{70} & $\{2,0,0,1\}$ & 70 & 42/5 \\
 \textbf{70'} & $\{0,0,0,4\}$ & 70 & 72/5 \\
 \textbf{75} & $\{0,1,1,0\}$ & 75 & 8 \\
 \textbf{105} & $\{0,0,1,2\}$ & 105 & 52/5 \\
 \textbf{126} & $\{2,0,1,0\}$ & 126 & 10 \\
 \textbf{126'} & $\{5,0,0,0\}$ & 126 & 20 \\
 \textbf{160} & $\{3,0,0,1\}$ & 160 & 63/5 \\
 \textbf{175} & $\{1,1,0,1\}$ & 175 & 48/5 \\
 \textbf{175'} & $\{1,2,0,0\}$ & 175 & 12 \\
 \textbf{175''} & $\{0,3,0,0\}$ & 175 & 72/5 \\
\\[-2.ex]
	\hline
	\end{tabular}
\vspace{0pt}
	\captionsetup{width=\linewidth}
	\captionof{table}{Lowest-dimensional $SU(5)$ irreducible representations and their Casimir invariants.}	
	\label{tab:su5_specs}
  \end{minipage}\hfill
  \begin{minipage}[b]{.5\textwidth}
    \centering
    \includegraphics[width=\linewidth]{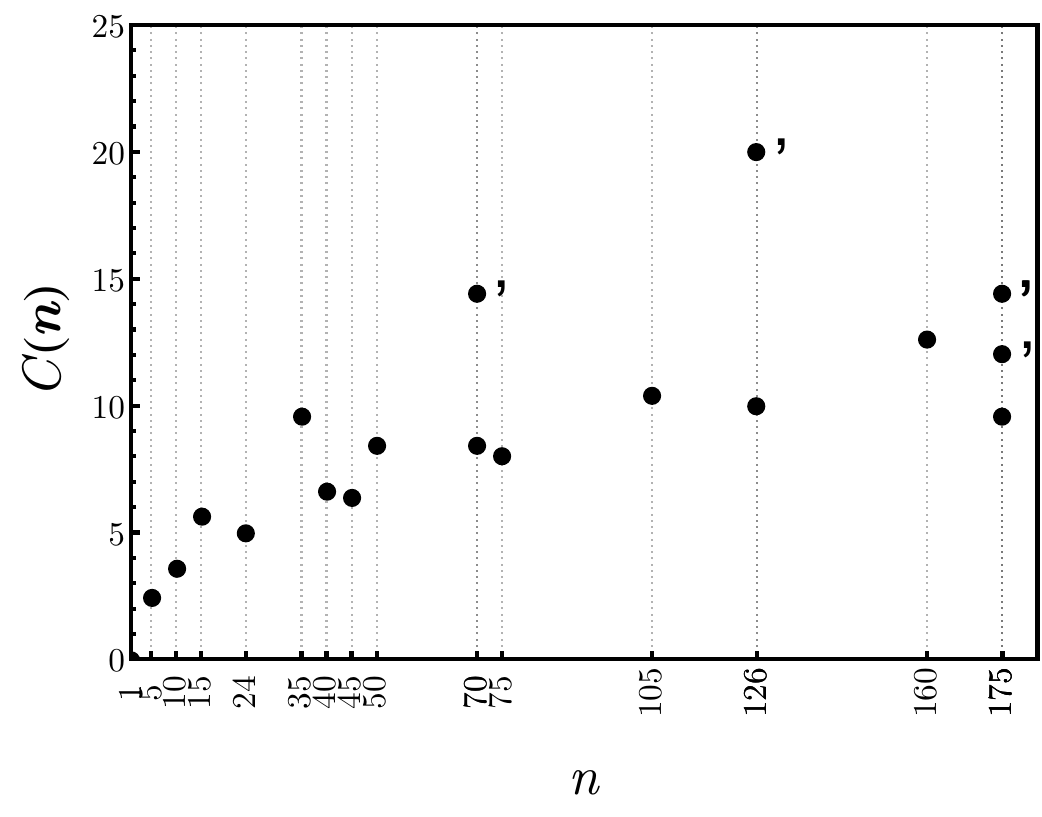}
    \captionof{figure}{$SU(5)$ Casimir invariant as a function of the dimension of the irreducible representation.}
\vspace{70pt}
    \label{fig:cn_su5}
  \end{minipage}\hfill
\end{figure}

The dimension of the adjoint representation is $\tilde{n}_{SU(5)} = 24$.
Table \ref{tab:su5_specs} shows the lowest-dimensional $SU(5)$ irreducible representations ($\textbf{\textit{n}}$), alongside their Dynkin indices, dimensions ($n$), and quadratic Casimir invariants ($C(\textbf{\textit{n}})$). In Figure \ref{fig:cn_su5} we plot the quadratic Casimir invariant as a function of the dimension of irreducible representations found in Table \ref{tab:su5_specs}. The primed (double primed) marker $\bullet\textbf{'}$ ($\bullet\textbf{''}$) represents a primed (double primed) representation $\textbf{\textit{n}'}$ ($\textbf{\textit{n}''}$).

\subsection{$SO(10)$ Gauge Group}

\begin{figure}[h!]
  \begin{minipage}[b]{.45\textwidth}
	\centering
	\begin{tabular}{ c c c c } 
	\hline
	irrep  & Dynkin & Dimension & Casimir  \\
	($\textbf{\textit{n}}$) & Index & ($n$) & ($C(\textbf{\textit{n}})$) \\
	\hline\hline
\\[-2.ex]
 \textbf{1} & $\{0,0,0,0,0\}$ & 1 & 0 \\
 \textbf{10} & $\{1,0,0,0,0\}$ & 10 & 9/2 \\
 \textbf{16} & $\{0,0,0,0,1\}$ & 16 & 45/8 \\
 \textbf{45} & $\{0,1,0,0,0\}$ & 45 & 8 \\
 \textbf{54} & $\{2,0,0,0,0\}$ & 54 & 10 \\
 \textbf{120} & $\{0,0,1,0,0\}$ & 120 & 21/2 \\
 \textbf{126} & $\{0,0,0,2,0\}$ & 126 & 25/2 \\
 \textbf{144} & $\{1,0,0,1,0\}$ & 144 & 85/8 \\
 \textbf{210} & $\{0,0,0,1,1\}$ & 210 & 12 \\
 \textbf{210'} & $\{3,0,0,0,0\}$ & 210 & 33/2 \\
\\[-2.ex]
	\hline
	\end{tabular}
	\captionsetup{width=\linewidth}
	\captionof{table}{Lowest-dimensional $SO(10)$ irreducible representations and their Casimir invariants.}	
\vspace{10pt}
	\label{tab:so10_specs}
  \end{minipage}\hfill
  \begin{minipage}[b]{.5\textwidth}
    \centering
    \includegraphics[width=\linewidth]{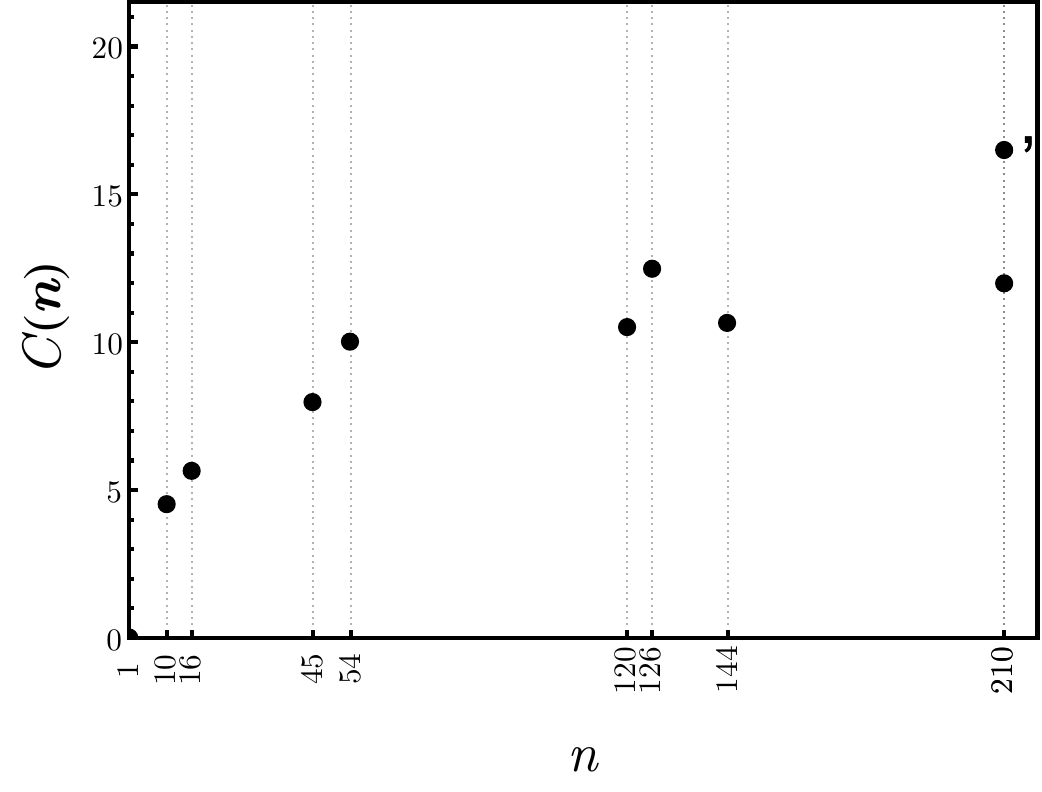}
    \captionof{figure}{$SO(10)$ Casimir invariant as a function of the dimension of the irreducible representation.}
\vspace{0pt}
    \label{fig:cn_so10}
  \end{minipage}\hfill
\end{figure}

The dimension of the adjoint representation is $\tilde{n}_{SO(10)} = 45$.
Table \ref{tab:so10_specs} shows the lowest-dimensional $SO(10)$ irreducible representations ($\textbf{\textit{n}}$), alongside their Dynkin indices, dimensions ($n$), and quadratic Casimir invariants ($C(\textbf{\textit{n}})$). In Figure \ref{fig:cn_so10} we plot the quadratic Casimir invariant as a function of the dimension of irreducible representations found in Table \ref{tab:so10_specs}. The primed marker $\bullet\textbf{'}$ represents a primed representation $\textbf{\textit{n}'}$.

\subsection{$E_6$ Gauge Group}

\begin{figure}[h!]
  \begin{minipage}[b]{.45\textwidth}
	\centering
	\begin{tabular}{ c c c c } 
	\hline
	irrep  & Dynkin & Dimension & Casimir  \\
	($\textbf{\textit{n}}$) & Index & ($n$) & ($C(\textbf{\textit{n}})$) \\
	\hline\hline
\\[-2.ex]
 \textbf{1} & $\{0,0,0,0,0,0\}$ & 1 & 0 \\
 \textbf{27} & $\{1,0,0,0,0,0\}$ & 27 & 26/3 \\
 \textbf{78} & $\{0,0,0,0,0,1\}$ & 78 & 12 \\
 \textbf{351} & $\{0,0,0,1,0,0\}$ & 351 & 50/3 \\
 \textbf{351'} & $\{0,0,0,0,2,0\}$ & 351 & 56/3 \\
 \textbf{650} & $\{1, 0, 0, 0, 1, 0\}$ & 650 & 18 \\
\\[-2.ex]
	\hline
	\end{tabular}
\vspace{38pt}
	\captionsetup{width=\linewidth}
	\captionof{table}{Lowest-dimensional $E_6$ irreducible representations and their Casimir invariants.}	
	\label{tab:e6_specs}
  \end{minipage}\hfill
  \begin{minipage}[b]{.5\textwidth}
    \centering
    \includegraphics[width=\linewidth]{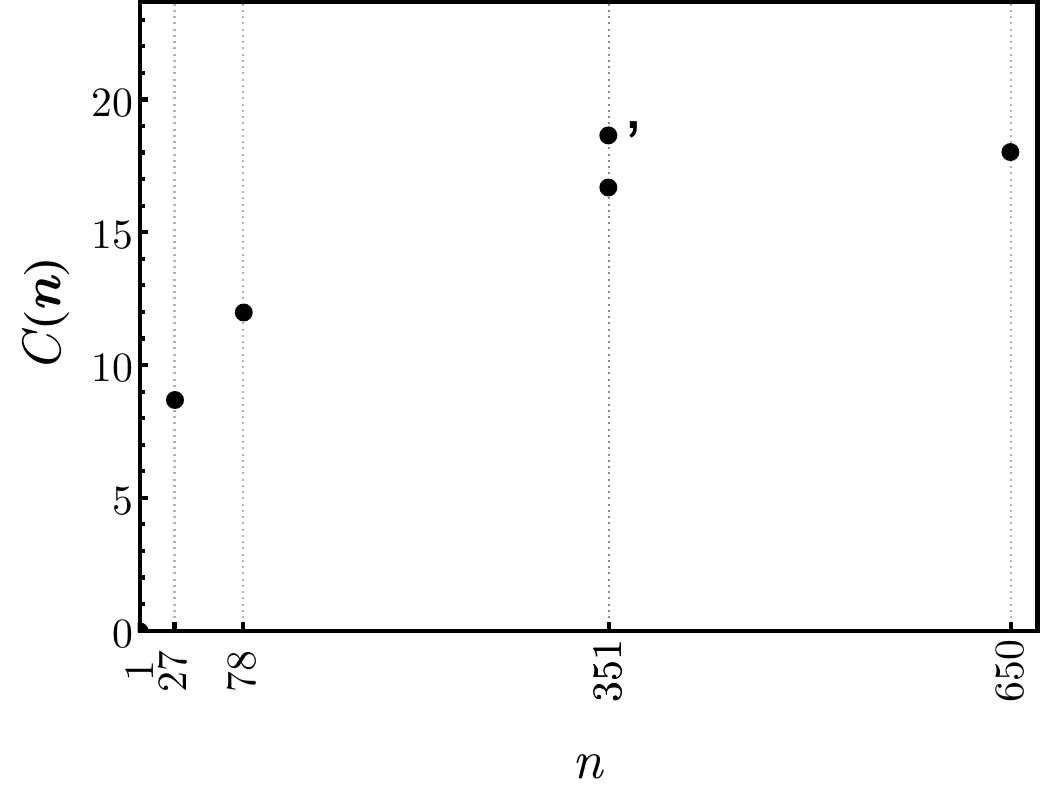}
    \captionof{figure}{$E_6$ Casimir invariant as a function of the dimension of the irreducible representation.}
\vspace{0pt}
    \label{fig:cn_e6}
  \end{minipage}\hfill
\end{figure}

The dimension of the adjoint representation is $\tilde{n}_{E_6} = 78$.
Table \ref{tab:e6_specs} shows the lowest-dimensional $E_6$ irreducible representations ($\textbf{\textit{n}}$), alongside their Dynkin indices, dimensions ($n$), and quadratic Casimir invariants ($C(\textbf{\textit{n}})$). In Figure \ref{fig:cn_e6} we plot the quadratic Casimir invariant as a function of the dimension of irreducible representations found in Table \ref{tab:e6_specs}. The primed marker $\bullet\textbf{'}$ represents a primed representation $\textbf{\textit{n}'}$.

\newpage

\vspace*{20pt}
\begin{spacing}{1}
\bibliographystyle{JHEP}
{\footnotesize\bibliography{unitarity_GUTs.bib}}
\end{spacing}

\end{document}